\documentclass[aps,prab,reprint,groupedaddress,nofootinbib,floatfix]{revtex4-2}

\usepackage{csquotes}
\usepackage{amsmath} 
\usepackage{amsfonts}
\usepackage{amssymb}
\usepackage{graphicx}
\usepackage{multirow}
\usepackage{hyperref}
\usepackage[separate-uncertainty = true,multi-part-units=single]{siunitx}

\usepackage{tikz}
\usetikzlibrary{arrows.meta}
\usetikzlibrary{shapes}
\tikzset{%
  >={Latex[width=2mm,length=2mm]},
            rect/.style = {rectangle, rounded corners, draw=black,
                           minimum width=4cm, minimum height=0.8cm,
                           text width=4cm,
                           text centered, font=\sffamily, fill=orange!15,
                           transform shape},
           diam/.style = {diamond, draw=black,
               aspect=1.8, 
               text centered, font=\sffamily, fill=blue!30,
               transform shape},
           red/.style = {rectangle, rounded corners, draw=black,
           minimum height=0.8cm, text centered, font=\sffamily, fill=red!30,
           minimum width=1.5cm, transform shape}
}

\usepackage[acronym]{glossaries}
\newacronym{std}{std}{standard deviation}
\newacronym[plural=MCs,firstplural=main cavities (MCs), longplural={main cavities}]{MC}{MC}{main cavity}
\newacronym[plural=HCs,firstplural=harmonic cavities (HCs), longplural={harmonic cavities}]{HC}{HC}{harmonic cavity}
\newacronym[plural=HOMs,firstplural=higher order modes (HOMs), longplural={higher order modes}]{HOM}{HOM}{higher order mode}
\newacronym{PTBL}{PTBL}{periodic transient beam loading}
\newacronym[plural=CBIs,firstplural=coupled-bunch instabilities (CBIs),longplural=coupled-bunch instabilities]{CBI}{CBI}{coupled-bunch instability}
\newacronym{IBS}{IBS}{intra-beam scattering}
\newacronym{NC}{NC}{normal conducting}
\newacronym{SC}{SC}{superconducting}
\newacronym{DFT}{DFT}{discrete Fourier transform}
\newacronym{FT}{FT}{Fourier transform}
\newacronym{DFB}{DFB}{direct rf feedback}
\newacronym{PI}{PI}{proportional-integral}
\newacronym{NFP}{NFP}{near flat potential}
\newacronym{FP}{FP}{flat potential}
\newacronym{ALBuMS}{\texttt{ALBuMS}}{Algorithms for Longitudinal MultiBunch Beam Stability}

\usepackage[english]{babel}
\usepackage{hyperref}
\begin{document}


\title{Semi-analytical algorithms to study longitudinal beam instabilities \\ in double rf systems}


\author{A. Gamelin}
\email[]{alexis.gamelin@synchrotron-soleil.fr}
\author{V. Gubaidulin}
\affiliation{Synchrotron SOLEIL, L'Orme des Merisiers, Saint-Aubin, France}

\author{M. B. Alves}
\affiliation{Brazilian Synchrotron Light Laboratory -- LNLS, Brazilian Center for Research in Energy and Materials -- CNPEM, 13083-970, Campinas, SP, Brazil.}
\affiliation{Gleb Wataghin Institute of Physics, University of Campinas -- UNICAMP, 13083-859, Campinas, SP, Brazil}

\author{T. Olsson}
\affiliation{Helmholtz-Zentrum Berlin für Materialien und Energie, 14109 Berlin, Germany}


\date{\today}

\begin{abstract}
    Double rf systems are critical for achieving the parameters of 4th-generation light sources. 
    These systems, comprising both main and harmonic rf cavities, relax statistical collective effects but also introduce instabilities, such as Robinson and periodic transient beam loading (PTBL) instabilities.
    In this paper, we provide semi-analytical algorithms designed to predict and analyze these instabilities with improved accuracy and robustness. 
    The algorithms leverage recent advancements in the field, offering a computationally efficient and accurate complement to multibunch tracking simulations. 
    Using the SOLEIL~II project as a case study, we demonstrate how these algorithms can optimize rf cavity parameters in high-dimensional parameter spaces, thereby maximizing the Touschek lifetime.
    An open-source Python package, \texttt{ALBuMS} (Algorithms for Longitudinal Multibunch Beam Stability), is provided as an accessible tool for double rf system stability analysis.
\end{abstract}

\maketitle

\section{Introduction}

Double rf systems, including both \glspl{MC} and \glspl{HC}, are becoming a de facto standard in storage-ring-based light sources. What was unusual in 2\textsuperscript{nd}-generation light sources, common in 3\textsuperscript{rd}-generation is now nearly mandatory with the advent of 4\textsuperscript{th}-generation synchrotron light sources. The advantages of such systems operated to flatten the rf potential have been demonstrated many times. The increased bunch length induced by the near flat potential increases the bunch volume density, which relaxes statistical collective effects~\cite{leemann_interplay_2014} such as Touschek and \gls{IBS}. This is especially important in current and future 4\textsuperscript{th}-generation synchrotron light sources in order to not spoil the ultra-low transverse emittances, which are reduced to hundreds or tens of \si{\pico \meter \radian}. Other harmful consequences of collective effects, such as the additional energy spread induced by microwave instability, can also be significantly reduced or mitigated by the increased bunch length or the Landau damping generated by the increased synchrotron tune spread \cite{mosnier1999microwave}.

However, \glspl{HC} come with their own set of longitudinal beam instabilities, which can prevent the achievement of large bunch lengthening factors. Double rf systems can drive \enquote{Robinson} instabilities \cite{bosch_robinson_2001}, generally defined as \glspl{CBI} of mode $\ell = 0$, for which all bunches oscillate in phase, driven by the fundamental mode of the cavities. Different kinds of Robinson instabilities exist, which can be driven by various mechanisms and can excite different azimuthal (i.e. synchrotron) modes $m$. The \gls{PTBL} instability, also called coupled-bunch mode $\ell = 1$ instability, causes the beam to fall into a quasi-stable state, slowly drifting along the bunches with a rather long time period \cite{venturini_passive_2018, he_periodic_2022}. Instabilities usually observed in storage rings with a single rf system, such as \gls{CBI} driven by cavity \glspl{HOM} \cite{bosch_instabilities_2005, tavares_beam-based_2022} or equilibrium phase instability \cite{MIYAHARA1987518}, are still a concern for double rf systems.

A very careful design of the double rf system is then needed to anticipate, characterize and possibly avoid such instabilities. The stability of double rf systems is often estimated using multibunch macro-particle tracking simulations where many effects (long-range wakes of \gls{MC} and \gls{HC}, broadband short-range wakes, non-uniform filling patterns, feedback loops, ...) can be taken into account together \cite{borland_elegant:_2000, gamelin_mbtrack2_2021}. Several open-source tracking codes are available \cite{mbtrack2, gamelin_mbtrack2_2021, borland_elegant:_2000, pyAT, blond, stable}. However, such simulations are very computationally heavy and time-consuming, often requiring dedicated computing clusters and days of computation for a single parameter set. In addition, while tracking simulations are usually more accurate for real user cases, they often fail to provide a clear physical understanding, unless backed up by analytical models to explain the results. Simulations with high computational load can be a setback in the early phase of a project, when \gls{MC} and \gls{HC} parameters are still under specification. These constraints can lead to a reduction of the sampled parameter space and to the choice of sub-optimal solutions.

In such cases, it is much more efficient to resort to theoretical and semi-analytical methods, which can be much faster than tracking simulation, to explore the parameter space. The first step in this direction is to compute the longitudinal equilibrium of the beam, by solving a system of coupled Ha\"{i}ssinski equations \cite{haissinski_exact_1973}. This step is crucial as it is a common basis needed for most theoretical analyses to estimate if the solution found might be stable or not. Furthermore, it is also a very good tool to have when doing tracking simulations as it gives a very fast estimate of what the simulation result is expected to be if the beam is stable. Therefore, it is not surprising that many recent developments have tackled this matter \cite{venturini_passive_2018, olsson_self-consistent_2018, warnock_equilibrium_2020, warnock_equilibrium_2021, warnock_equilibrium_2021-1, he_longitudinal_2021, gamelin_equilibrium_2021, alves_equilibrium_2023}. The coupled Ha\"{i}ssinski equations can now be resolved with nonuniform filling patterns \cite{warnock_equilibrium_2020, warnock_equilibrium_2021, warnock_equilibrium_2021-1, he_longitudinal_2021}, multiple passive or active cavities \cite{gamelin_equilibrium_2021} and all the previous features with the addition of any arbitrary impedance \cite{alves_equilibrium_2023}. A few open-source implementations of these methods are available \cite{borland_elegant:_2000, mbtrack2, pycolleff, BunchLengthening_He}.

Once the beam equilibrium is found, the usual next step is using it to evaluate a given instability. But to reach a similar level as in tracking simulation, i.e. to be able to replicate the driving mechanism of a large variety of instabilities, it is important to evaluate the overall stability of the Ha\"{i}ssinski solution with respect to different instability mechanisms. For example, this kind of algorithm was developed by Bosch~et~al. in a series of papers \cite{bosch_suppression_1993, bosch_robinson_2001, bosch_instabilities_2005}.

In this article, we propose a modernized algorithm based on the Bosch one. The main modifications make use of the advances which the field has seen since the Bosch study was conducted. The modifications allow for better convergence for the equilibrium calculation and improved instability predictions. Moreover, our approach considers an additional instability, the \gls{PTBL}, and a different method to compute Robinson instabilities. The \gls{PTBL} is considered in the algorithm using two different models. The limitations of the new algorithm are discussed, and this method is shown to be in good agreement when benchmarked against multibunch tracking simulation using \texttt{mbtrack2} \cite{gamelin_mbtrack2_2021}. Then, using the context of the SOLEIL~II project, we show how this method can be used to do large parameter scans and find optimal cavity parameters to maximize Touschek lifetime. Finally, we show how cavity and machine impedance short-range wakes can be partially included in the new algorithm. Their effects on the fast mode coupling instability are studied for the first time both using tracking and this semi-analytical method.

In addition to this paper, a Python package named \gls{ALBuMS} is provided as an open-source tool based on the semi-analytic methods explored here to study the longitudinal beam stability in double rf systems. Such a package can be helpful to guide cavity design, tracking simulations and experimental measurement campaigns \cite{albums}.

\section{Algorithm for beam stability}
\label{sec:algo}

In this section, the principles of the original Bosch algorithm are presented, its limitations and the modifications done to achieve an improved algorithm are discussed. The main concepts regarding double rf systems and the notations used are presented in Appendix \ref{sec:double_rf_sys}. Through the article, the quantities with index 1 refer to the \gls{MC}, while index 2 refers to the \gls{HC}.

\subsection{Bosch original algorithm}
\label{subsec:bosch}
\begin{figure*}
\begin{minipage}{0.49\textwidth}
    \centering
    \begin{tikzpicture}[node distance=1.4cm,
    every node/.style={fill=white, font=\sffamily}, align=center, scale=1.0]
  \node (A)     [rect]              {Compute synchronous phase};
  \node (B)     [rect, below of=A]  {Apply optimal MC detuning};
  \node (C)     [rect, below of=B]   {Taylor expansion \\ of rf potential};
  \node (D)     [rect, below of=C]   {Compute bunch length};
  \node (E)     [rect, below of=D]   {Compute bunch form factors};
  \node (F)     [diam, below of=E, node distance=1.7cm]    {Converged ?};
  \node (G)     [rect, below of=F, node distance=2.3cm] {CBI driven by HOMs};
  \node (H)     [rect, below of=G] {Zero frequency instabilities};
  \node (I)     [rect, below of=H] {Robinson instabilities w/ or w/o mode coupling};
  \node (J)     [rect, below of=I] {CBI of mode $\ell \pm 1$};
  \draw[->]             (A) -- (B);
  \draw[->]             (B) -- (C);
  \draw[->]             (C) -- (D);
  \draw[->]             (D) -- (E);
  \draw[->]             (E) -- (F);
  \draw[->]       (F.east) to[bend right=90] node {No} (A.east);
  \draw[->]             (F) -- node {Yes} (G);
  \draw[->]             (G) -- (H);
  \draw[->]             (H) -- (I);
  \draw[->]             (I) -- (J);

  \end{tikzpicture}
      \end{minipage}
      \begin{minipage}{0.49\textwidth}
    \centering
    \begin{tikzpicture}[node distance=1.5cm,
    every node/.style={fill=white, font=\sffamily}, align=center, scale=1.0]
  \node (A)     [rect]              {Set MC \& HC parameters};
  \node (B)     [rect, below of=A]  {Solve Ha\"{i}ssinski equation};
  \node (conv1)  [diam, below of=B, node distance=1.7cm]    {Converged ?};
  \node (conv1a) [red, right of=conv1, xshift=2.5cm]    {Stop};
  \node (C)      [rect, below of=conv1, node distance=2.5cm]   {Taylor expansion \\ of rf potential};
  \node (D)     [rect, below of=C]   {CBI driven by HOMs};
  \node (E)      [rect, below of=D]    {Zero frequency instabilities};
  \node (F)      [rect, below of=E] {Robinson instabilities w/ or w/o mode coupling};
  \node (G)       [rect, below of=F] {PTBL instability};
  \draw[->]             (A) -- (B);
  \draw[->]             (B) -- (conv1);
  \draw[->]             (conv1) -- node {No} (conv1a);
  \draw[->]             (conv1) -- node {Yes} (C);
  \draw[->]             (C) -- (D);
  \draw[->]             (D) -- (E);
  \draw[->]             (E) -- (F);
  \draw[->]             (F) -- (G);
  \end{tikzpicture}
        \end{minipage}

    \caption{Bosch algorithm (left) and modified algorithm (right) for the study of longitudinal beam instabilities in double rf systems with passive harmonic cavities.}
    \label{fig:diag_bosch_passive}
\end{figure*}

The original Bosch algorithm from \cite{bosch_robinson_2001} used to estimate beam stability for a passive \gls{HC} is described in the left diagram of Fig.~\ref{fig:diag_bosch_passive}. The main input to the algorithm is the lattice parameters and the rf system description. Particularly for a passive \gls{HC}, the main input is its tuning angle $\psi_{2}$, which controls the \gls{HC} voltage.

The first part of the algorithm is a convergence loop to solve the usual self-consistent problem in which bunch distribution and beam-induced voltage are interdependent. Based on the input parameters, the synchronous phase is computed using the energy balance conditions, taking into account the additional losses induced by the passive \gls{HC}. The effect of the bunch profile is accounted for by using initial values for the scalar bunch form factor for both the \gls{MC} and \gls{HC} \cite{wilson1994fundamental}. Secondly, the \gls{MC} tuning is set to the optimal tuning condition, Eq.~\eqref{eq:optimal_tuning}. In the third step, the rf effective potential $U(\tau)$ is expanded in Taylor series up to the fourth order using Eqs.~(4-6) of \cite{bosch_robinson_2001}. In this method, coefficients of this Taylor expansion are used to characterize the shape of the rf potential created by the double rf system. These coefficients are then used to compute the bunch length by integrating this potential, Eq.~(8) of \cite{bosch_robinson_2001}. The choice of integration boundaries can be an important parameter for the Bosch algorithm, as presented in Appendix \ref{sec:convergence}. From the computed bunch length, new values for the scalar form factors are calculated for both the \gls{MC} and \gls{HC}. If the obtained form factors are significantly different from the initial ones, these steps are repeated with new initial values for the form factors, based on a weighted average of the new and previous values \cite{bosch_suppression_1993, bosch_robinson_2001}, until convergence is achieved.

Then, using the converged results from the first part of the algorithm, different types of instabilities are considered successively. Bosch analysis of Robinson instabilities is based on the linearized Vlasov equation for a quadratic potential (corresponding to Gaussian bunches and a single rf system) \cite{wang_modes_1987, krinsky_longitudinal_nodate} where the linear synchrotron frequency $\omega_s$ is replaced by the coherent dipole oscillation frequency $\omega_R$ of the double rf system. This coherent dipole oscillation frequency is computed by approximating the dipole motion as a small rigid oscillation for a symmetric bunch and given by Eq.~(10) of \cite{bosch_robinson_2001}.

The first type of instability considered is longitudinal \gls{CBI} driven by \glspl{HOM}. The expression used for the complex frequency shift corresponds to the usual \gls{CBI} one \cite{tavares_beam-based_2022}, applied at the resonance of the \gls{HOM}, where the synchrotron frequency is replaced by $\omega_R$. The computed growth rate is compared with the synchrotron damping rate and with an estimated Landau damping rate based on the Taylor expansion coefficient computed previously \cite{bosch_suppression_1993, bosch_robinson_2001}. If the resulting growth rate is still positive, the beam is predicted to be unstable due to a \gls{CBI} driven by \glspl{HOM}.

The instability analysis continues with the zero-frequency instabilities of different azimuthal $m$ modes, Eq.~(14) of \cite{bosch_robinson_2001}. The dipole zero-frequency instability corresponds to the usual equilibrium phase instability \cite{robinson1964stability, MIYAHARA1987518}, also sometimes called static Robinson instability \cite{PhysRevAccelBeams.26.044401}. This occurs when the beam-induced voltage component moving with the beam overcomes the restoring force provided by the generator voltage. The higher-order ($m>1$) zero-frequency instabilities correspond to the same instability mechanism but for higher-order perturbations (i.e. bunch length for quadrupole zero-frequency instability, and so on). If, for each $m$ mode, the zero-frequency instability has not been predicted to be unstable, then the algorithm continues to its next steps. 

The equation system for the Robinson instability (i.e. coupled bunch mode $\ell = 0$) corresponding to a given azimuthal mode $m$ (Eq.~(13) and (15) of \cite{bosch_robinson_2001}) is solved iteratively to get the corresponding complex coherent angular frequency $\Omega$, from which the radiation damping has already been subtracted. If mode coupling is considered between the dipole and quadrupole modes, the equations for the coupled system are used (Eqs.~(B11) and (B13) of \cite{bosch_robinson_2001}). If the solution corresponds to instability growth, the Landau damping rate estimated for azimuthal mode $m$ based on the Taylor expansion of the rf potential is subtracted from the growth rate to evaluate if the Landau damping effect can stabilize the beam. If an instability is predicted, it is named dipole Robinson instability if $m=1$, quadrupole Robinson instability if $m=2$, and so on\footnote{In Ref.~\cite{bosch_robinson_2001}, a distinction is made between the Robinson instabilities computed with the coupled or uncoupled system. For the example of mode $m=1$, the name \enquote{coupled-dipole Robinson instability} is used for the case with mode coupling and \enquote{dipole Robinson instability} for the case without. This distinction is not made in this article as all results show the coupled system results.}. The \enquote{fast mode coupling instability} refers to the instability driven by the merging of the dipole and quadrupole modes.

Finally, the \gls{CBI} for coupled-bunch mode $\ell = \pm 1$ is considered by solving the same equation system based on the linearized Vlasov equation for a quadratic potential as for the uncoupled dipole ($m=1$) instability but evaluated for the coupled bunch mode $\ell = \pm 1$ instead of $\ell = 0$.
Landau damping is then considered in the same manner as for \gls{CBI} driven by \gls{HOM} to tell if \gls{CBI} for mode $\ell = \pm 1$ is predicted.

For active \gls{HC}, the algorithm is largely similar to the one for passive \gls{HC}. Only the first part of the algorithm is slightly different. The $\xi$ value, i.e. the ratio of the harmonic to main cavity \enquote{force} as defined in Eq.\eqref{eq:xi_bosch}, is treated as an input parameter to the algorithm to compute cavity phases $\theta_1$ and $\theta_2$. Then, the Taylor expansion of the rf potential is done, followed by bunch length and form factor calculation, as is done for the passive case. Both \gls{MC} and \gls{HC} are set to optimal detuning, taking into account the form factor previously computed. These steps correspond to the same ones as shown in left side of Fig.~\ref{fig:diag_bosch_passive} but are executed in a different order and without the convergence loop. Then, the second part regarding instability estimation is kept the same.

\subsection{Bosch algorithm limitations}
\label{sec:algo_lim}

There are several approximations which are done along the different steps of the Bosch algorithm, which may place limitations on its predictive power and practical usage. 

The first issue we found is that the convergence loop used to solve the self-consistent problem can be quite sensitive to the integration boundaries used to compute the bunch length in the passive \gls{HC} algorithm described in left diagram of Fig.~\ref{fig:diag_bosch_passive}. In Appendix \ref{sec:convergence}, we show that, for some lattice parameters, it leads to large errors in the instability prediction. In addition, for active \gls{HC}, the form factors and associated parameters are not obtained self-consistently, which is a possible error source.

The Landau damping estimate compares the instability growth rate to the synchrotron frequency spread multiplied by a constant factor, which depends on the rf potential Taylor expansion and the instability synchrotron mode \cite{bosch_suppression_1993, bosch_robinson_2001}. This approximation does not consider if the incoherent synchrotron frequency spread overlaps with the instability coherent frequency, which is a necessary condition for Landau damping. A more rigorous approach would be to compute the stability diagram to check that the complex coherent frequency is indeed within the Landau contour to assess stability, similar to \cite{venturini_passive_2018, PhysRevAccelBeams.24.011002}. Even then, when multiple azimuthal modes are relevant to the instability, the standard approach using a stability diagram also fails.

The instability estimate for the \gls{CBI} mode $\ell = \pm 1$ uses the same formula as the uncoupled dipole Robinson instability applied to $\ell = \pm 1$. This instability has since then been studied in depth, analytically \cite{venturini_passive_2018}, using tracking codes \cite{he_periodic_2022} under the name \acrfull{PTBL} instability and, more recently, also experimentally \cite{PhysRevAccelBeams.27.044403}. It was shown that this instability is quite peculiar as it is mainly driven by the reactive (imaginary) part of the impedance and as the instability manifests itself by slowly drifting bunch profile variations in the bunch index space. Recent findings suggest that the mechanism driving this instability might depend on several $m$ modes \cite{venturini_passive_2018, alves2024} and $\ell$ modes \cite{gamelin_beam_2022}. Also, as the instability coherent frequency is particularly low, Landau damping cannot play a role if there is no overlap with the synchrotron frequency spread \cite{alves2024}. This insight could explain why the prediction for \gls{CBI} mode $l = \pm 1$ using the original Bosch algorithm may not be accurate.

Additionally, the analysis in \cite{bosch_robinson_2001} is restricted to symmetric bunches: scalar form factors are used, and the underlying theory assumes a quadratic potential. In all cases, the models used for instability prediction are only strictly valid for uniform filling patterns. Also, the model is expected to give good predictions for $\xi \leq 1$, but for $\xi > 1$, its accuracy may be reduced because the bunch may have a double hump shape.

\subsection{Modified algorithm}
\label{subsec:modif_algo}

\newcommand{\minitab}[2][l]{\begin{tabular}{#1}#2\end{tabular}}

\begin{table*}[t]
\centering
\caption{Longitudinal instabilities considered by the modified algorithm and their corresponding mode numbers.}
\begin{ruledtabular}
    \begin{tabular}{c|cccccc}\hline
    \multirow[c]{2}{*}{Instability name} & Dipole & Quadrupole & Fast mode & Zero  & Periodic Transient  & Coupled bunch \\
    & Robinson & Robinson  & coupling & frequency & Beam Loading & driven by \glspl{HOM} \\
    \hline
    \multirow[c]{3}{*}{\minitab[c]{Other names used \\ in the literature}} & \multirow[c]{3}{*}{AC Robinson} & \multirow[c]{3}{*}{-} & \multirow[c]{3}{*}{-} & \multirow[c]{3}{*}{\minitab[c]{Equilibrium phase\footnote{Only for the dipole mode ($m=1$).}, \\ Static Robinson\footnotemark[1], \\ DC Robinson\footnotemark[1]}} & \multirow[c]{3}{*}{Mode $\ell=1$} & \multirow[c]{3}{*}{-} \\
    & &  & & &  & \\
    & &  & & &  & \\
    \hline
    \multirow[c]{2}{*}{\minitab[c]{Coupled bunch \\ mode number}} & \multirow[c]{2}{*}{$\ell=0$} & \multirow[c]{2}{*}{$\ell=0$} & \multirow[c]{2}{*}{$\ell=0$} & \multirow[c]{2}{*}{$\ell=0$} & \multirow[c]{2}{*}{$\ell= 1\footnote{$\ell=1$ is the dominant mode.},2,3,...$} & \multirow[c]{2}{*}{\minitab[c]{Depends on the \\ driving \gls{HOM}}} \\
    & &  & & &  & \\
    \hline
    \multirow[c]{3}{*}{\minitab[c]{Azimuthal (synchrotron) \\ mode number}} & \multirow[c]{3}{*}{$m=1$} & \multirow[c]{3}{*}{$m=2$} & \multirow[c]{3}{*}{\minitab[c]{$m=1$ \\ $m=2$ \\ (coupled)}} & \multirow[c]{3}{*}{\minitab[c]{$m=1$, \\ or $m=2$, \\ or $m=3$, ...}} & \multirow[c]{3}{*}{$m= 1,2,3,...$} & \multirow[c]{3}{*}{$m=1$} \\
    & &  & & &  & \\
    & &  & & &  & \\
    \hline
    \multirow[c]{3}{*}{\minitab[c]{Landau damping \\ estimate from Taylor \\ expansion of rf potential}} & \multirow[c]{3}{*}{Yes} & \multirow[c]{3}{*}{Yes} & \multirow[c]{3}{*}{Yes} & \multirow[c]{3}{*}{No} & \multirow[c]{3}{*}{No} & \multirow[c]{3}{*}{Yes} \\
    & &  & & &  & \\
    & &  & & &  & \\
    
    \end{tabular}
\end{ruledtabular}
\label{tab:instabilities}

\end{table*}

To answer some of these limitations (regarding convergence issues, asymmetric bunches and the \gls{PTBL} instability), we propose a modified algorithm working for both passive and active \glspl{HC} as shown in right diagram of Fig.~\ref{fig:diag_bosch_passive}. The input values are now the cavity voltage, phase and detuning for active cavities and the detuning for passive cavities. Compared to the original algorithm, it allows studying beam stability with the cavities in non-optimal tuning conditions where the instabilities may behave differently.


Then, the self-consistent loop to get the scalar form factors is replaced by computing the full beam equilibrium by solving a system of coupled Ha\"{i}ssinski equations \cite{haissinski_exact_1973}. At this level, this modification allows us to take into account asymmetric bunch profiles correctly. When the full longitudinal bunch distributions are known, the information needed for the next steps can be extracted: \gls{std} bunch length, complex form factors and beam-induced voltage. Two different open-source implementations of this kind of \enquote{Ha\"{i}ssinski solver} are used. The first one, here called \enquote{Venturini} solver, is implemented in the \texttt{BeamLoadingEquilibrium} class of \texttt{mbtrack2}~\cite{mbtrack2} which is based on the method developed by Venturini \cite{venturini_passive_2018} for uniform filling patterns, later modified to work also with active cavities \cite{gamelin_equilibrium_2021}. The second one, here called \enquote{Alves} solver, is implemented in the \texttt{LongitudinalEquilibrium} class of \texttt{pycolleff} \cite{pycolleff}, which allows finding the beam equilibrium for arbitrary filling patterns both in time and frequency domain, and including machine impedance \cite{alves_equilibrium_2023}. As shown in Appendix~\ref{sec:convergence}, compared to the self-consistent part of the original algorithm, solving the Ha\"{i}ssinski equation using these methods is much more robust against numerical parameter variations. Also, unlike the original algorithm, which self-consistent loop seems to always converge toward a form factor value, the algorithm stops if no solution to the Ha\"{i}ssinski equation can be found. 

The Taylor expansion of the rf potential used to get the Landau damping estimate is done in the same way as in the Bosch algorithm, based on the solution found at the previous step. Next, the instability estimate for the \gls{CBI} driven by \glspl{HOM} and zero frequency instabilities follow the original algorithm. The equation system to solve Robinson instabilities with or without mode coupling is written in a general way for an arbitrary number of cavities. As the convergence of the complex coherent frequency can sometimes be difficult to achieve, both Powell's and Broyden's methods are used to solve the equation system by using their \texttt{scipy} implementation \cite{10.1093/comjnl/7.2.155, broyden1965class}.

The instability analysis for \gls{CBI} of mode $\pm 1$ is replaced by two different models depending on which method is used to solve the Ha\"{i}ssinski equation.

If the \enquote{Venturini} solver is used, the \gls{PTBL} instability criterion derived by He in \cite{he_novel_2022} is used to estimate the beam stability. This instability criterion is based on a phase perturbation, following the coupled bunch mode $\ell=1$ pattern, applied to the equilibrium solution. If the resulting phase perturbation is larger than the initial one, the \gls{PTBL} instability is predicted. In most cases, this model is enough to predict the \gls{PTBL}, but it was shown that it could be inaccurate for some cases like the MAX~IV one \cite{PhysRevAccelBeams.27.044403} where the instability also depends on azimuthal mode $m=2$ \cite{harmonLIP_alves_2024} and a model that only considers phase perturbations cannot capture the relevant instability mechanism.

If the \enquote{Alves} solver is used instead, the new algorithm uses the model presented in \cite{alves2024, alves2024theoretical}, referred to as Gaussian longitudinal mode coupling instability (Gaussian LMCI) theory, and implemented in \texttt{pycolleff} \cite{pycolleff}. This model is based on the linearized Fokker-Planck equation for a quadratic potential but uses the formalism of Suzuki's longitudinal mode coupling theory for Gaussian bunches \cite{suzuki1982mode}. It allows coupling between different azimuthal and radial modes, and an effective incoherent synchrotron frequency $\overline{\omega_{s,q}}$, based on the longitudinal matching condition, is used to account for the effect of the \gls{HC}:~$\overline{\omega_{s,q}} = |\eta| \sigma_{\delta} \slash \sigma_{\tau}$,
where $\eta$ is the slip factor, $\sigma_{\delta}$ is the relative energy spread and $\sigma_{\tau}$ is the \gls{std} bunch length. In this way, the Gaussian LMCI model is more general than Bosch's model and He's phase perturbation, as it is possible to account for the interaction of several azimuthal modes of any coupled-bunch mode of interest, while the other models are specialized to specific modes. Although Fokker-Planck terms could be included in the Gaussian LMCI calculations, in our studies we noted that neglecting these terms does not influence the agreement with tracking results, therefore all the reported results were obtained with a Vlasov-based Gaussian LMCI model for simplicity. 
In both cases, the Landau damping is neglected, which should be a reasonable approximation due to the low coherent frequency signature of \gls{PTBL} as discussed in Sec.~\ref{sec:algo_lim}.

If all the instability calculations have converged and none of them are predicted, the beam stability is inferred. If no solution to the Ha\"{i}ssinski equation can be found or if an instability calculation fails, the algorithm result is a \enquote{not converged}\footnote{In most cases, non-convergence of the numerical calculations cannot be used to infer information about the stability of the solution, other than being an indication of strong collective effects. Thus the different classification in the algorithm.} flag in \gls{ALBuMS}.

Even though, as shown in Secs.~\ref{sec:robinson} and~\ref{subsec:non_uniform}, the Gaussian LMCI method can also be used to compute Robinson instabilities, it is not how the Robinson instabilities are solved by default as will be discussed later in Sec.~\ref{sec:discussion}. Unless directly specified, the Bosch equations described in subsections \ref{subsec:bosch} and \ref{subsec:modif_algo} are used. However, it is implemented as an optional possibility in \gls{ALBuMS} for the \enquote{Alves} solver.

Throughout the paper, we will refer to each algorithm by the name of the used \enquote{Ha\"{i}ssinski solver}: \enquote{Bosch} solver for the original algorithm, \enquote{Venturini} and \enquote{Alves} solvers for the modified algorithm using either the He criterion or the Gaussian LMCI for predicting the \gls{PTBL}. Table \ref{tab:instabilities} presents a summary of the different longitudinal instabilities considered by the modified algorithm.

\section{Algorithm comparison and benchmarks}

This section aims to illustrate the different algorithms described in Sec.~\ref{sec:algo}. The different test cases are shown for the SOLEIL~II storage ring, focusing on \gls{NC} passive type \glspl{HC}. We start, in Sec.~\ref{sec:tracking}, by presenting the tracking setup used. Then, in Sec.~\ref{sec:robinson}, the algorithms are benchmarked with tracking simulations for the particular case of the fast mode coupling instability and, in Sec.~\ref{sec:res}, for the general SOLEIL~II case. Finally, the two \gls{PTBL} models are compared in Sec.~\ref{sec:ptbl}. The benchmark of the \gls{ALBuMS} implementation of the original Bosch algorithm and a convergence study of an important numerical parameter for the \enquote{Bosch} solver is shown in Appendix \ref{sec:convergence}.

The SOLEIL~II project aims to upgrade the SOLEIL storage ring and injectors, the French 3\textsuperscript{rd}-generation light source, to a 4\textsuperscript{th}-generation light source \cite{susini2024brief, nadji2023upgrade, noauthor_soleil_nodate}. The storage ring parameters~\cite{loulergue:ipac2024-tupg47} used for this study are reported in Table \ref{tab:param}.
\begin{table}
        \centering
        \caption{Parameters used for SOLEIL~II storage ring in this work. The shunt impedance $R_s$ is given using the $V^2/(2P)$ definition where $V$ and $P$ are the total voltage and power in the cavity.}
        \begin{ruledtabular}
        \begin{tabular}{ l c } 
        {Parameter} & {Value}\\
        \hline
         Electron beam energy $E_0$ & \SI{2.75}{\giga \electronvolt} \\ 
         Natural emittance & \SI{84}{\pico \metre} \\
         Circumference $L$ & \SI{353.97}{\metre} \\ 
         Revolution frequency $f_{0}$ & \SI{846.9}{\kilo \hertz} \\
         Rf frequency $f_{{\mathrm{rf}}}$ & \SI{352.3}{\mega \hertz} \\
         Harmonic number $h$ & 416 \\
         Momentum compaction factor $\alpha_c$ & \num{1.057e-04} \\
         Energy loss per turn (w/o IDs) $U_0$ & \SI{469}{\kilo \electronvolt} \\
         Longitudinal damping time $\tau_L$ & \SI{11.64}{\milli \second} \\
         Relative energy spread $\sigma_\delta$ & \num{9.06e-4} \\
         Natural bunch length $\sigma_\tau$ & \SI{8.9}{\pico \second} \\ 
         Main rf voltage $V_1$ & \SI{1.7}{\mega \volt} \\ 
         Main cavity (total) shunt impedance $R_{s_1}$ & \SI{20}{\mega \ohm} \\
         Main cavity quality factor $Q_{0_1}$ & \num{35.7e3} \\
         Main cavity loaded quality factor $Q_{L_1}$ & \num{6e3} \\
        \end{tabular}
        \end{ruledtabular}
    \label{tab:param}
\end{table}

\subsection{Tracking setup}
\label{sec:tracking}

The beam dynamics simulations, used to benchmark the modified algorithms, are obtained using the tracking functionalities of the \texttt{mbtrack2} Python collective effect library \cite{gamelin_mbtrack2_2021}. \texttt{mbtrack2} can simulate the multiple bunches of a beam in a parallelized way using the \texttt{mpi4py} library \cite{9439927}. The tracking model is a simple longitudinal and transverse one-turn map in which different non-linear effects are taken into account (radiation effects, higher order momentum compaction factor, chromaticity and amplitude-dependent tune shifts). Both single-bunch and multibunch collective effects are taken into account. Both the active \gls{MC} and passive \gls{HC} are simulated using \texttt{CavityResonator} class objects \cite{yamamoto_investigation_2019}, which compute both generator and beam loading voltage applied to the beam. An example script of \texttt{mbtrack2} simulations used in this work is provided in Supplemental Material \cite{supp}.

The \gls{MC} total phase $\theta_1$ is set to approximately cancel the losses from the \gls{NC} passive \gls{HC} as described in Appendix~\ref{subsec:MC_settings}. Then the \gls{MC} tuning angle $\psi_1$ is set to the optimal tuning condition for the beam current $I_0$, and finally, the initial generator parameters are computed. The same tuning as for the tracking simulations is applied to the \gls{MC} for the algorithm using the modified methods. The \gls{MC} tuning set when using the \enquote{Bosch} method is, however, slightly different as the \gls{MC} tuning is part of the self-consistent loop and takes the scalar form factor into account.

The \gls{MC} total cavity voltage and phase are controlled in the simulation using a \gls{PI} loop \cite{yamamoto_reduction_2018}. The \gls{PI} loop is set up to have a very low proportional gain and mostly works with the integral gain. If the proportional gain is set too high, the \gls{PI} loop itself can trigger a dipole Robinson or a fast mode coupling instability if it is close to its threshold. The loop delay and the averaging period for the cavity phasor monitoring are set to \SI{2}{\micro \second} and \SI{0.6}{\micro \second} respectively, which are typical values for this type of rf loop.

To better match with the theory in the benchmark, higher orders of momentum compaction factor of the SOLEIL~II lattice are excluded in the simulations (even if no strong effects were observed when it was added). A uniform filling of the storage ring is assumed, and each of the 416 bunches is represented by \num{10000} macro-particles\footnote{This number may seem small but is actually enough to accurately get the beam voltage from the \gls{MC} and \gls{HC} considering their low-frequency impedance.}. The beam is tracked for \num{300000} turns, corresponding approximately to 30 times the longitudinal radiation damping time, to make sure the \gls{MC} voltage and phase have converged to their set points and to allow the detection of instabilities with low growth rates.

Table \ref{tab:speed} indicates the execution speed of the different methods. Both \enquote{Venturini} and \enquote{Alves} solvers are only somewhat slower than the \enquote{Bosch} one (respectively \num{2.4} times and \num{19} times slower) compared to the significant gain in accuracy. Compared to the other solvers, the \enquote{Alves} solver can be used for non-uniform fill patterns by the cost of increasing its execution speed by a factor \num{100}. All these solvers are orders of magnitude faster than tracking, even the one parallelized with a bunch per core (i.e. 416 cores), as done here. In practice, running large parameter scans with thousands of parameter combinations and optimization methods with tracking is out of reach even when using supercomputers but becomes possible with semi-analytical methods.
\begin{table}
    \centering
    \caption{Execution speed for different methods to evaluate a single parameter set.}
        \begin{ruledtabular}
            \begin{tabular}{ l r }
                {Method} & {Time per evaluation} \\
                \hline
                 \enquote{Bosch} solver & \SI[multi-part-units=brackets]{1.22(0.03)e-2}{\second} \\
                 \enquote{Venturini} solver & \SI[multi-part-units=brackets]{2.95(0.53)e-2}{\second} \\
                 \enquote{Alves} solver (uniform fill) & \SI[multi-part-units=brackets]{2.33(0.13)e-1}{\second} \\
                 \enquote{Alves} solver (non-uniform fill) & \SI[multi-part-units=brackets]{2.94(0.23)e1}{\second} \\
                \texttt{mbtrack2} tracking & \SI[multi-part-units=brackets]{3.38(0.1)e4}{\second} \\
            \end{tabular}
        \end{ruledtabular}
    \label{tab:speed}
\end{table}

\subsection{Fast mode coupling instability}
\label{sec:robinson}

\begin{figure}
    \centering
    \includegraphics[width=1.0\columnwidth]{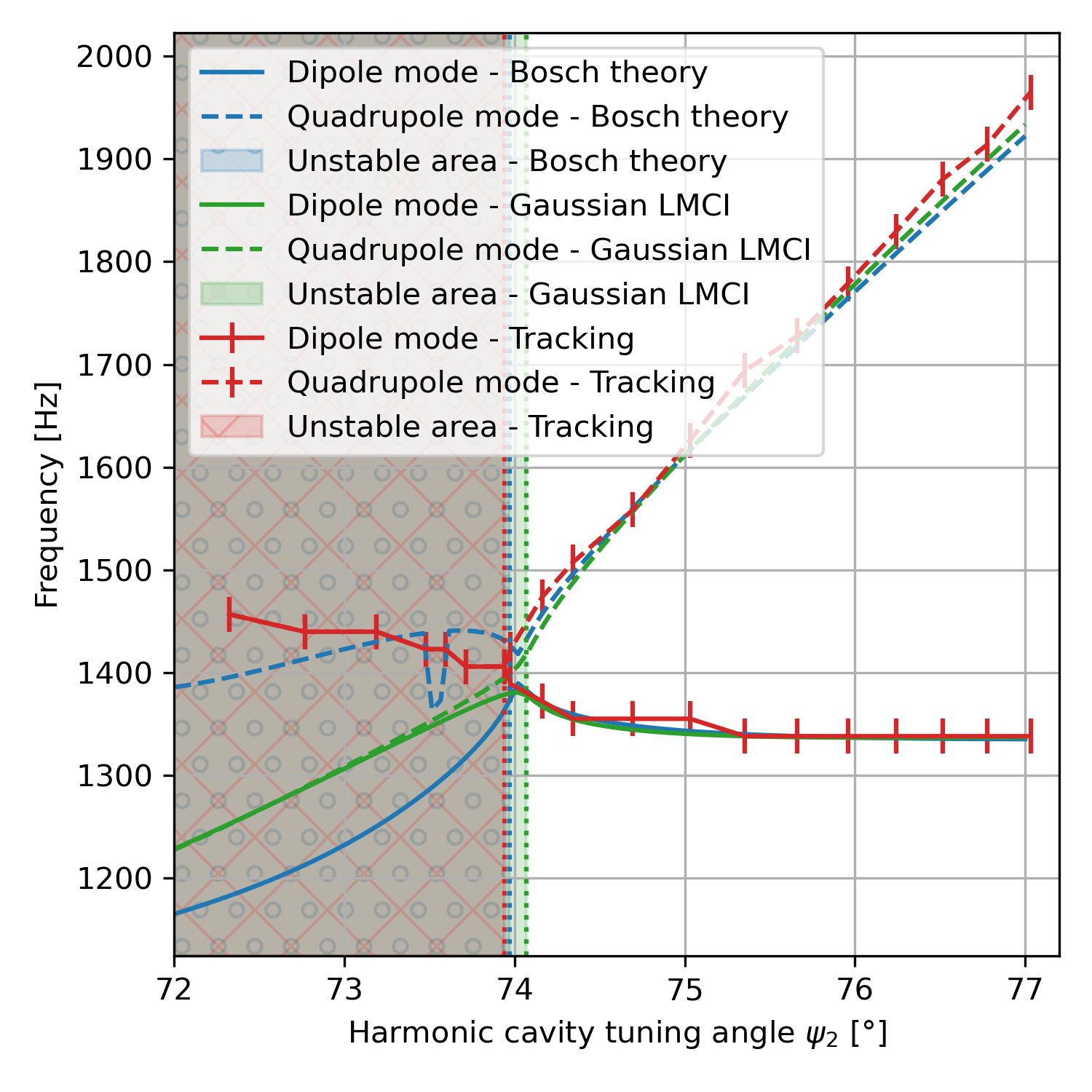}
    \caption{Dipole and quadrupole mode frequencies as a function of the \gls{HC} tuning angle $\psi_2$ in SOLEIL~II storage ring at \SI{200}{\milli \ampere} beam current. Passive \gls{HC} parameters: $R_{s_2}/Q_{0_2}=\SI{90}{\ohm}$, $Q_{0_2} =\num{36e3}$.}
    \label{fig:mode_coupling}
\end{figure}

A benchmark for the fast mode coupling instability between analytic predictions and tracking results is shown in Fig.~\ref{fig:mode_coupling}. 
The analytic prediction from Bosch's theory of mode coupling is obtained using the \enquote{Venturini} solver. Both \enquote{Bosch} and \enquote{Alves} solvers (using Bosch theory) give similar results with some small numerical discrepancies. Additionally, the analytic prediction using the Gaussian LMCI theory is also shown.

In tracking, a specific setting is used to observe these modes.
The longitudinal motion is excited by a frequency sweep around the synchrotron tune to excite the dipolar oscillation mode.
The dipole mode is obtained by recording the \enquote{dipolar coherent spectrum} (i.e. the \gls{DFT} of bunch center of mass). This process is similar to simulating a synchrotron tune measurement in tracking. The quadrupole mode could be observed in the usual simulation, even without longitudinal excitation, but instead looking at the \enquote{quadrupolar coherent spectrum} (i.e. the \gls{DFT} of bunch length). The error bars shown for the tracking data correspond to the \gls{DFT} resolution using \num{50000} turns. 

Both the Bosch theory (here computed using the \enquote{Venturini} solver) and Gaussian LMCI theory match very well with tracking results both for the computed mode frequencies and for the predicted threshold of the fast mode coupling instability resulting from the merging of both dipole and quadrupole modes. When both modes couple and the instability is triggered, strong center of mass and bunch length oscillations are observed in tracking with amplitudes which depend on the instability strength. Typical oscillation amplitude for the center of mass is \SI{50}{\pico \second} peak-to-peak and for the \gls{std} bunch length \SI{15}{\pico \second} peak-to-peak. All the bunches oscillate in phase, corresponding to coupled-bunch mode $\ell=0$ as expected for Robinson instabilities. For this particular case, the fast mode coupling instability would limit the \gls{HC} tuning to a maximum of $\xi \approx 0.84$, far from the \gls{NFP}, see Appendix \ref{sec:nfp}.

\subsection{Application to SOLEIL~II storage ring}
\label{sec:res}

\begin{figure*}
    \centering
    \includegraphics[width=1.0\textwidth]{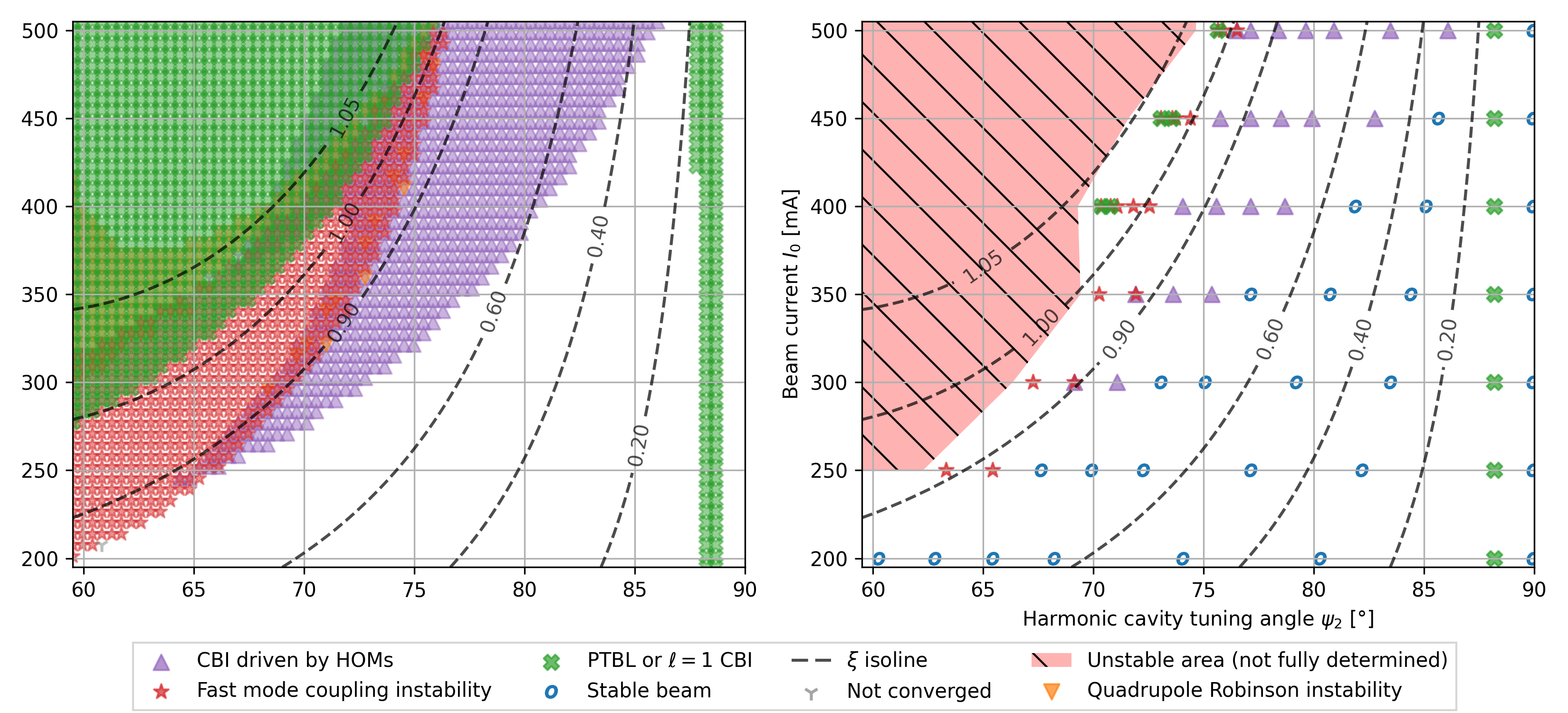}
    \caption{Instabilities predicted for the SOLEIL~II storage ring, with the \enquote{Alves} solver (left) and with tracking (right). A \gls{HOM} is included with $f_\text{HOM} = \SI{1.7}{\giga \hertz}$, $Q_{0_\text{HOM}}=670$ and $R_{s_\text{HOM}}=\SI{8.8}{\kilo \ohm}$. Passive \gls{HC} parameters: $R_{s_2}/Q_{0_2}=\SI{60}{\ohm}$, $Q_{0_2} = \num{31e3}$.}
    \label{fig:HOM_Alves}
\end{figure*}

\begin{figure}
    \centering
    \includegraphics[width=1.0\columnwidth]{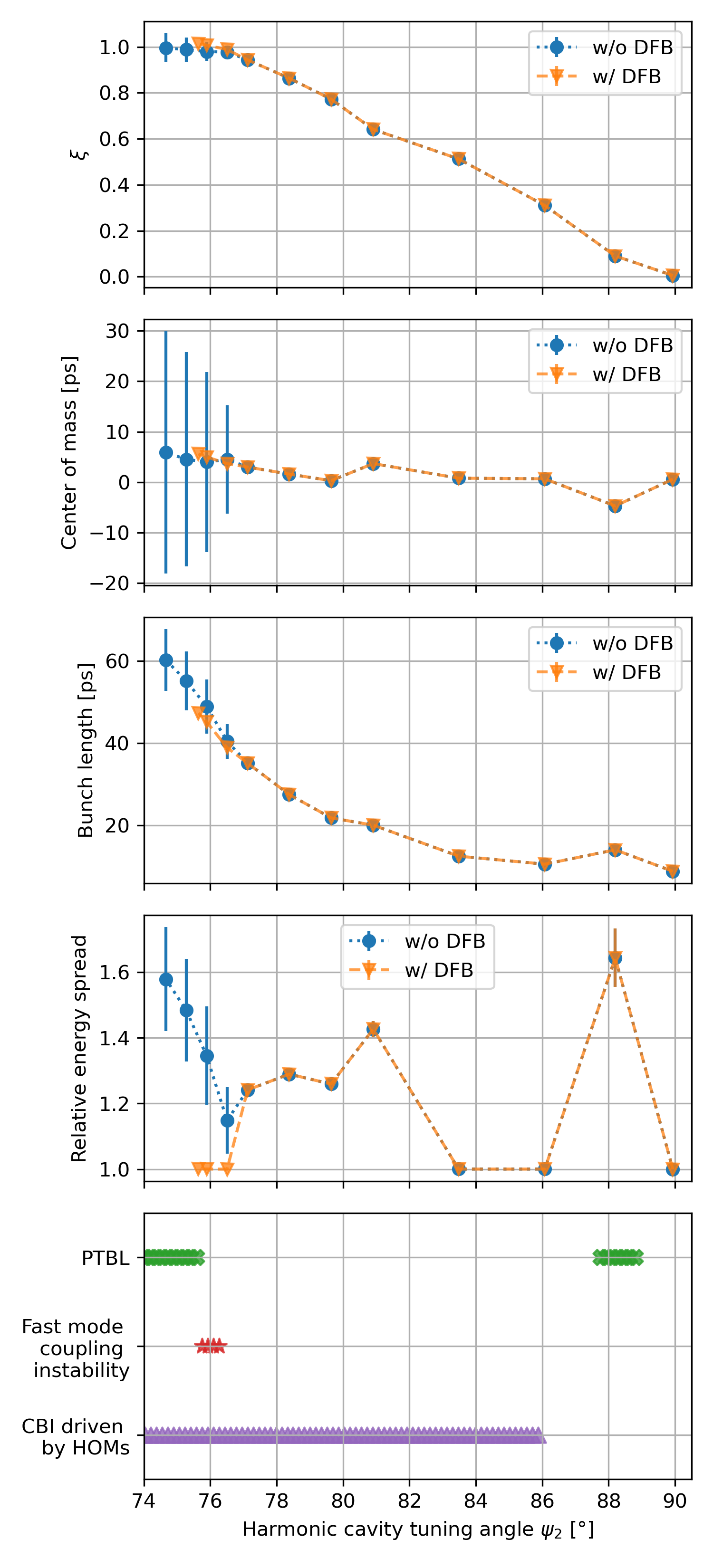}
    \caption{Tracking results (first four top plots) for the SOLEIL~II lattice at \SI{500}{\milli \ampere} beam current with and without \gls{DFB}. Dots and triangles show the mean value over the last \num{1000} turns while error bars show the standard deviation of the value over the last \num{1000} turns. Instability prediction using the \enquote{Alves} solver (bottom). Passive \gls{HC} parameters: $R_{s_2}/Q_{0_2}=$ \SI{60}{\ohm}, $Q_{0_2} =$ \num{31e3}.}
    \label{fig:Tracking}
\end{figure}

\begin{figure}
    \centering
    \includegraphics[width=1.0\columnwidth]{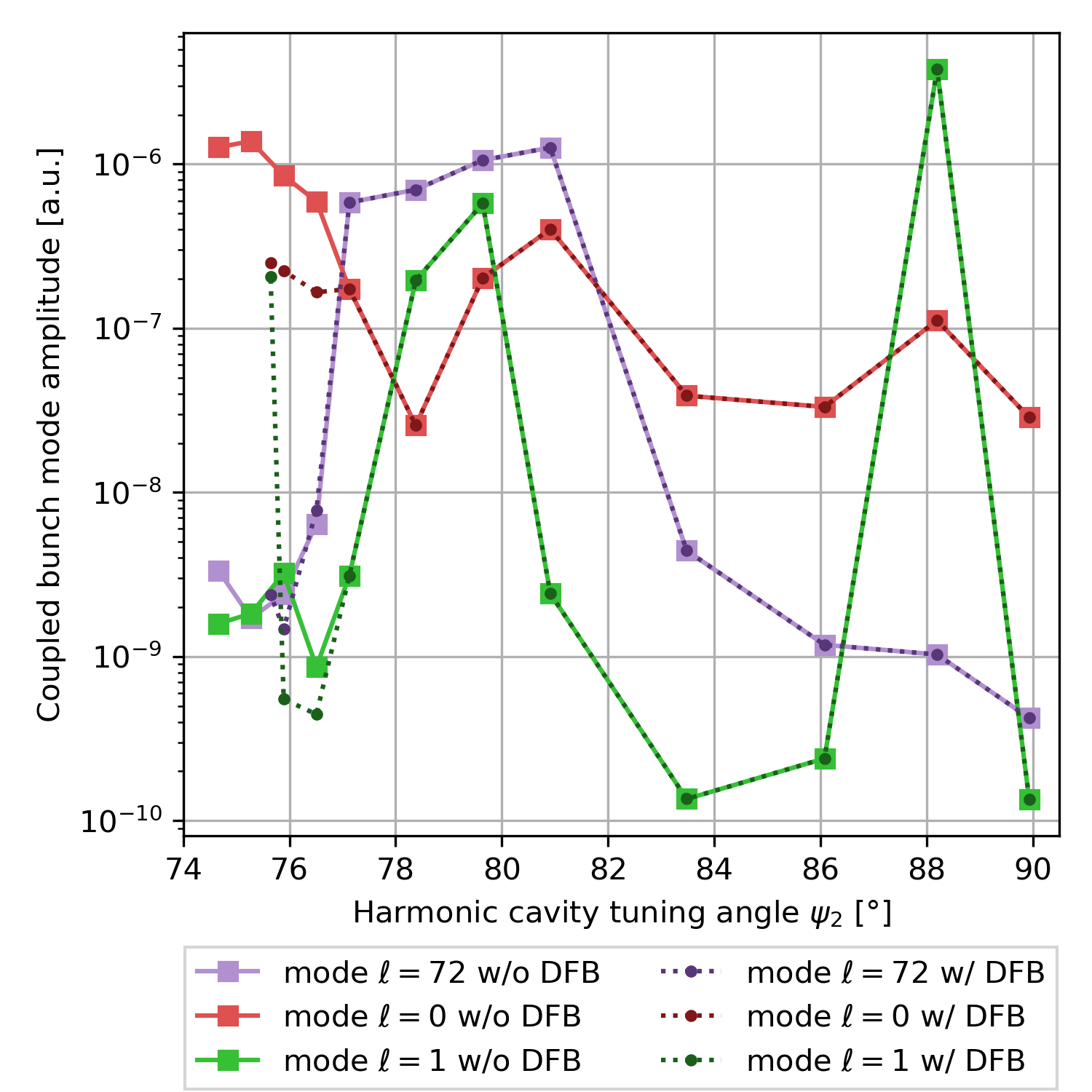}
    \caption{Coupled bunch mode $\ell$ amplitude computed from tracking results for the SOLEIL~II lattice at \SI{500}{\milli \ampere} beam current with and without \gls{DFB}. Passive \gls{HC} parameters: $R_{s_2}/Q_{0_2}=\SI{60}{\ohm}$, $Q_{0_2} =\num{31e3}$.}
    \label{fig:Tracking_CBM}
\end{figure}

In this section, we present the benchmark of the modified algorithm full usage against tracking simulation for the SOLEIL~II storage ring. A single \gls{HOM}, corresponding to the strongest \gls{HOM} from the impedance computation of 4 \glspl{MC}, is included in the benchmark. The general view is shown in Fig.~\ref{fig:HOM_Alves} where large scans of both beam current $I_0$ and \gls{HC} tuning angle $\psi_2$ are done. In Fig.~\ref{fig:Tracking} and Fig.~\ref{fig:Tracking_CBM}, the results are analyzed in depth for \SI{500}{\milli \ampere} beam current.

On the left side of Fig.~\ref{fig:HOM_Alves}, the full instability prediction with the \enquote{Alves} solver is shown. For each pair of beam current $I_0$ and \gls{HC} tuning angle $\psi_2$ in the plot, the algorithm is used to compute the beam stability at the point. If the corresponding point is unstable, a marker related to the instability is shown. The instability map is overlapped with $\xi$ isolines, Eq.~\eqref{eq:xi_bosch}, showing how far that point is from the \gls{NFP} conditions.

On the right side of Fig.~\ref{fig:HOM_Alves}, the same kind of map is obtained using tracking simulations (tracking details are presented below). In that case, the blue circle markers specify the tracked positions for which the beam was found to be stable. The $\xi$ isolines shown on the right plot are obtained from the Ha\"{i}ssinski solution of the \enquote{Alves} solver (and thus are identical to the left plot ones). The $\xi$ values from tracking (not shown here) agree very well with semi-analytical ones in the regions where the beam is stable (or when stabilized by feedback).

Each of the points on the map corresponds to one or several tracking simulations, which were analyzed to determine if the beam was stable or not. If not stable, the triggered instability is identified by checking the amplitude of each coupled-bunch mode, the energy spread, the bunch profile evolution and the oscillations of the center of mass and bunch length. If the instability identified is the fast mode coupling, another tracking simulation is started in which the fast mode coupling is suppressed. This step is needed as the fast mode coupling instability can either prevent other instabilities from developing or trigger the \gls{HOM} coupled-bunch mode. If the required feedback gain to suppress the instability is too strong, it leads to a fast beam loss, and it is not possible to fully identify all the instabilities which could be triggered for this point on the map. A red hashed area is shown to specify the locations of the unstable yet not fully determined tracking points.

As an example of this analysis, the tracking results at \SI{500}{\milli \ampere} beam current are shown in Figs.~\ref{fig:Tracking} and~\ref{fig:Tracking_CBM}. This particular current value corresponds to the nominal beam current for the SOLEIL~II storage ring in uniform fill mode and is interesting because various instabilities are predicted.

The \gls{HOM} impedance is included in tracking as a \texttt{CavityResonator} object with a resonance frequency set at the closest beam harmonic, corresponding to the coupled bunch mode $\ell=344$: $f_\text{HOM} = (4 h + 344 + \nu_{s_0}) f_0 \approx$ \SI{1.7}{\giga \hertz}, where $\nu_{s_0} = 0.002$ is the incoherent synchrotron tune without HC. The shunt impedance and quality factor used for the \gls{HOM} are: $R_{s_\text{HOM}}=$ \SI{8.8}{\kilo \ohm} and $Q_{0_\text{HOM}}=670$. 

The coupled bunch instability driven by the \gls{HOM} is triggered when the \gls{HC} is tuned in, as the mean incoherent synchrotron frequency decreases and the \gls{CBI} growth rate increases beyond the longitudinal radiation damping rate.

In tracking, the coupled bunch mode spectrum is computed from the beam center of mass data every \num{50000} turns. If a single pick-up is used, such a spectrum reaches from coupled bunch mode $\ell=0$ to $\ell = h/2-1 = 207$, then the expected signal from coupled bunch mode $\ell=344$ can be observed as mode $\ell = h - 344 = 72$\footnote{More precise analysis could follow \cite{prabhakar_phase_1999}}. The amplitudes of coupled bunch modes $\ell =$ 0, 1 and 72 are shown in Fig.~\ref{fig:Tracking_CBM} for the last \num{50000} tracking turns.

When the \gls{HC} is fully detuned, i.e. $\psi_2 =$ \SI{90}{\degree} corresponding to $V_2=$ \SI{0}{\volt}, the beam is fully stable and the coupled bunch mode amplitudes recorded in Fig.~\ref{fig:Tracking_CBM} gives a reference point. At $\psi_2 =$ \SI{88}{\degree} a strong mode $\ell=1$ coupled bunch instability is observed, corresponding to the parked cavity case discussed in section~\ref{sec:ptbl}, as shown by the high amplitude of $\ell=1$ mode. 

When the \gls{HC} starts to be tuned in (between $\psi_2 =$ \SI{86}{\degree} and $\psi_2 =$ \SI{77}{\degree}) the strength of the \gls{CBI} driven by the \gls{HOM} increases as visible from the amplitude of mode $\ell=72$ and from the increased relative energy spread. It is visible in Fig. \ref{fig:Tracking_CBM} that from $\psi_2 =$ \SIrange{78}{80}{\degree}, the amplitude of mode $\ell=1$ is increased. This is a consequence of the excited \gls{CBI} driven by the \gls{HOM}, which raise the amplitude of most modes during the instability. The same simulation done without the \gls{HOM}, so without the excited \gls{CBI}, leads to a mode $\ell=1$ amplitude lower than \num{e-9}.

Near $\psi_2 =$ \SI{77}{\degree}, the fast mode coupling instability is triggered as shown by the center of mass and bunch length oscillations (where the error bars represent their amplitude). The amplitude increase of the coupled bunch mode $\ell=0$ is also a clear sign of a Robinson type instability. 

In the region close to $\psi_2 = \SI{76}{\degree}$, it is visible that the amplitude of coupled bunch mode $\ell=72$ drops. Here, the synchrotron tune spread provided by the \gls{HC} at the \gls{NFP} ($\xi \approx 1$) is stabilizing the \gls{CBI} driven by the \gls{HOM} through the Landau damping effect. However, this stabilization is not predicted correctly in the semi-analytical algorithm as a \gls{CBI} is expected for these values. 

Interestingly, the Landau damping effect seems correctly accounted for when $I_0 <$ \SI{400}{\milli \ampere} and $\xi \approx 1$, as shown by the good agreement in that region in Fig.~\ref{fig:HOM_Alves}. As described in Sec.~\ref{sec:algo}, the Landau damping rate computed in the semi-analytical method depends on the Taylor expansion made of the rf potential. When $I_0 \geq$ \SI{400}{\milli \ampere}, the bunch profile is rather asymmetric and the 3\textsuperscript{rd} and 4\textsuperscript{th} powers of the Taylor expansion have similar amplitudes, which reduce the estimated synchrotron frequency spread in this model (Eq. 17 of \cite{bosch_robinson_2001}). It is not the case when $I_0 <$ \SI{400}{\milli \ampere}, where the bunch shape tends toward double-bump shape and for which the Taylor expansion is dominated by the 3\textsuperscript{rd} power.

The overall behavior of longitudinal \gls{CBI} described here: an increased growth rate induced by the reduction of the mean incoherent synchrotron frequency, then damped via Landau damping, is expected when an \gls{HC} is included. This is in good agreement with the behavior reported in \cite{lindberg_theory_2018}, addressing both Landau damping and quartic potential more accurately.

For \gls{HC} tuning below $\psi_2 =$ \SI{77}{\degree}, the \gls{PTBL} instability is predicted analytically but not observed in tracking because of the large fluctuations induced by the fast mode coupling instability. To be able to observe the \gls{PTBL} instability in tracking, a \acrfull{DFB} \cite{akai_stability_2022, yamamoto_stability_2023}, which has the effect of reducing the effective shunt impedance seen by the beam, is applied to the \gls{MC} in order to suppress the fast mode coupling instability. The \gls{DFB} gain is set to a relatively low value corresponding to a reduction of the effective shunt impedance from $R_{s_1} = $ \SI{20}{\mega \ohm} to \SI{17.3}{\mega \ohm}. The \gls{DFB} delay and monitoring period values are set to be the same ones as for the \gls{PI} loop. As shown in Fig.~\ref{fig:Tracking}, the \gls{DFB} allows to stabilize the fast mode coupling instability and to stably reach the \gls{PTBL} threshold characterized by the rise of mode $\ell=1$ in Fig.~\ref{fig:Tracking_CBM}. In tracking, the \gls{PTBL} signature of the bunch profile shape drifting in the bunch index space is also observed.

For the tuning points close to $\psi_2 =$ \SI{75}{\degree}, the \gls{DFB} was not able to stabilize the fast mode coupling instability. If the \gls{DFB} gain is set to stronger values, a fast beam loss is observed. However, most likely, if the \gls{DFB} would be able to stabilize the beam, the \gls{PTBL} instability would be observed as its threshold was shown in tracking to be close to $\psi_2 =$ \SI{76}{\degree}. 

Overall, the agreement between the predicted instability using the semi-analytical algorithm and what is observed in tracking is excellent. Only a minor divergence in Landau damping strength is found near $\xi \approx 1$ for some bunch shapes obtained with high currents above \SI{400}{\milli\ampere}.

When the \gls{DFB} is not used, the threshold found by tracking for the fast mode coupling instability agrees well with the prediction. The agreement with the \enquote{Venturini} solver is also very good, as it gives quite similar results to the \enquote{Alves} solver. As shown in Fig.~\ref{fig:soleil_convergence} of Appendix~\ref{sec:convergence}, the small difference is that it predicts a slightly reduced fast mode coupling instability area near \SI{500}{\milli \ampere}, and that the \gls{PTBL} threshold is a bit more pessimistic. On the contrary, the original algorithm does not give a correct result. The \enquote{Bosch} solver predicts a stable beam, and neither the fast mode coupling instability nor the \gls{PTBL} are predicted in the \SI{500}{\milli \ampere} region as shown in Fig.~\ref{fig:soleil_convergence} of Appendix~\ref{sec:convergence}.

\subsection{PTBL models}
\label{sec:ptbl}

\begin{figure}
    \centering
    \includegraphics[width=1.0\columnwidth]{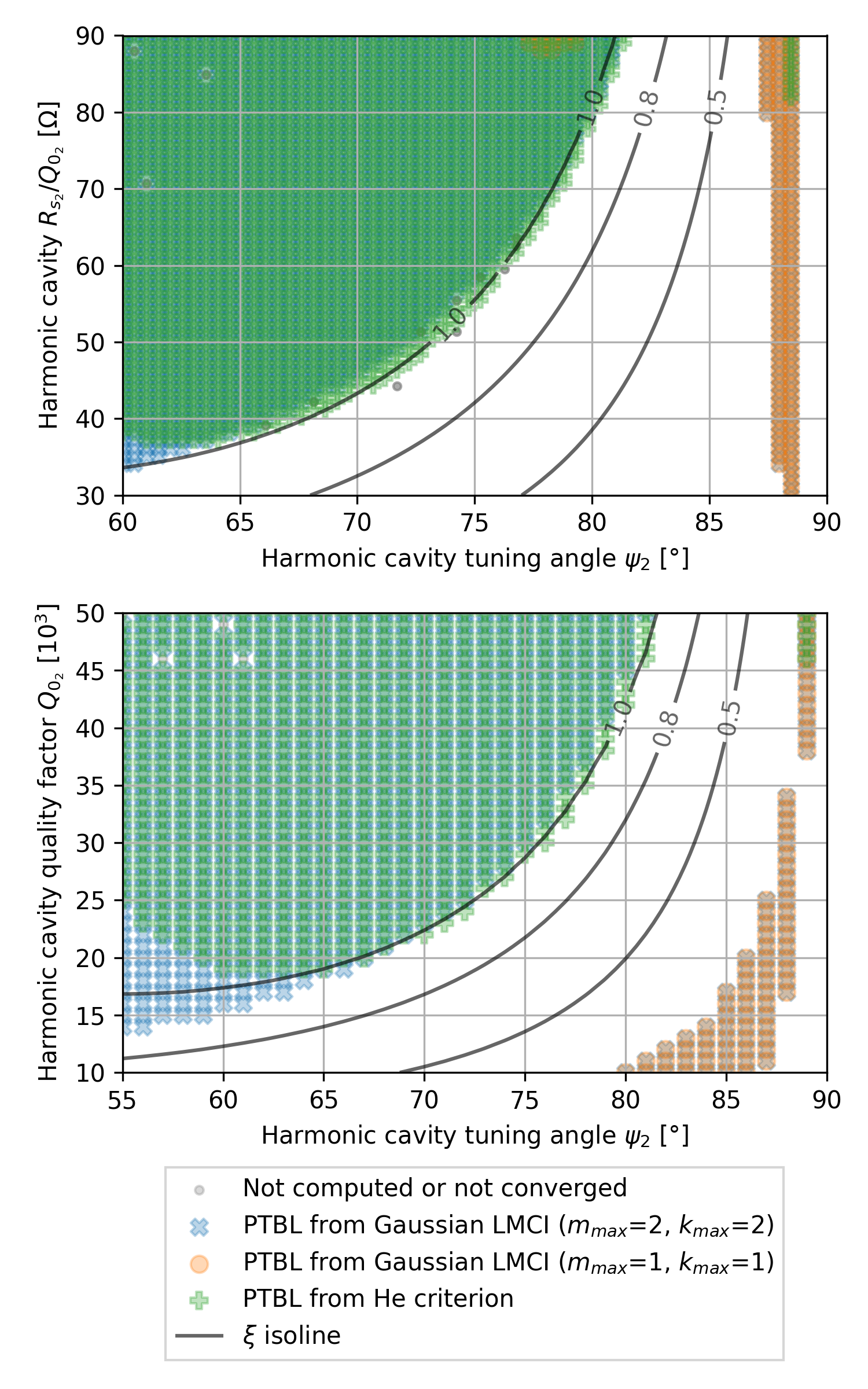}
    \caption{\gls*{PTBL} instability predicted by various models. Top: $R_{s_2}/Q_{0_2}$ is varied for different \gls{HC} tuning angle $\psi_2$ for a constant $Q_{0_2} =$ \num{31e3}. Bottom: $Q_{0_2}$ is varied for different \gls{HC} tuning angle $\psi_2$ for a constant $R_{s_2}/Q_{0_2}=$ \SI{60}{\ohm}.}
    \label{fig:compare_PTBL}
\end{figure}

The two models used for the \gls{PTBL} instability prediction are compared over a large range of \gls{HC} parameters in Fig.~\ref{fig:compare_PTBL}. In this section only, both the He criterion and the Gaussian LMCI method are computed using the \enquote{Alves} solver to get a fair comparison between both \gls{PTBL} models (as the \enquote{Venturini} solver does not work well for $\xi > 1.0$, see Appendix \ref{sec:convergence}).

The He criterion \cite{he_novel_2022} consistently predicts the \gls{PTBL} instability either when the $R_{s_2}/Q_{0_2} \geq$ \SI{40}{\ohm} or when $Q_{0_2} \geq$ \num{20e3}. When below these values, the He criterion predicts a stable beam. In contrast, \gls{PTBL} is predicted by the Gaussian LMCI method. This finding is consistent with \cite{he_novel_2022}, where the He criterion is shown to work well with \gls{SC} passive \gls{HC} type parameters (so high values of $Q_{0_2}$) and with \cite{PhysRevAccelBeams.27.044403}, in which the He criterion is shown to have a large error compared to measured instability for MAX~IV with its passive \gls{NC} \gls{HC} with $Q_{0_2}=$ \num{20.8e3}. 

When both models are predicting the instability, the He criterion is found to be a bit more pessimistic than the Gaussian LMCI, i.e., the threshold is found for the He criterion for a slightly larger \gls{HC} tuning angle $\psi_2$.

For the Gaussian LMCI, two cases are distinguished, when the azimuthal mode $m$ and radial mode $k$ up to $1$ ($m_{max} = 1$ and $k_{max}=1$) or two ($m_{max} = 2$ and $k_{max}=2$) are included. The results shows that the $m=2$ mode is critical to determine the \gls{PTBL} instability for parameters that are typical for \gls{NC} passive \glspl{HC}, suggesting why an analysis based on phase perturbation like the He criterion might not be accurate in those cases. Increasing $m_{max}$ and $k_{max}$ further, for example, up to the default value used in the rest of this paper, $m_{max} = 10$ and $k_{max}=10$, does not change the \gls{PTBL} instability prediction from the result found for $m_{max} = 2$ and $k_{max}=2$.

In both plots, the \gls{PTBL} instability is also predicted around \gls{HC} tuning angle $\psi_2 \approx$ \SI{88}{\degree}, nearly independently of the values of the $R_{s_2}/Q_{0_2}$ or $Q_{0_2}$. For such \gls{HC} tuning angle, corresponding to parking conditions where the \gls{HC} voltage is low, an instability is predicted for the He criterion for high $R_{s_2}/Q_{0_2}$ or high $Q_{0_2}$ and for nearly the full range for the Gaussian LMCI. This instability corresponds to a usual \gls{CBI}, as in single rf systems (as the \gls{HC} voltage is low) but driven by the parked \gls{HC} impedance. Using the standard coupled bunch formula \cite{[{For example see Eq. (51) pp. 139 of }]doi:10.1142/9789812778413_0009} applied to the \gls{HC} impedance (with $R_{s_2}/Q_{0_2}=$ \SI{60}{\ohm}, $Q_{0_2} =$ \num{31e3} and $\psi_2 =$ \SI{88}{\degree}), the largest growth rate is predicted for the coupled bunch mode $\ell = 1$ with \SI{160}{\per\second}, which is much larger than the synchrotron radiation damping rate. This translate the fact that for a detuned \gls{HC}, its resonance frequency is much closer to the mode $\ell = 1$ frequency $\nu f_{\mathrm{rf}} + f_{0}$ than for the real PTBL case when the \gls{HC} is tuned. The fact that the Gaussian LMCI method, and to a lesser degree the He criterion, also predicts this usual coupled bunch mode $\ell = 1$ instability is not a surprise as they are more general.

\section{Application to optimization problems}

The proposed algorithms can be used during rf system and cavity design to do large parameter scans in a considerably shorter time frame than by using tracking simulations. Taking advantage of the very high speed of these types of algorithms compared to tracking, we used them to find optimal \gls{HC} parameters for the SOLEIL~II storage ring to maximize Touschek lifetime.

The Touschek lifetime increase can be quantified by the ratio $R$ between the Touschek lifetime when using an \gls{HC} and the Touschek lifetime computed at the natural bunch length. Neglecting any changes due to the \gls{HC} in the energy acceptance and in the beam equilibrium parameters other than the bunch profile, this ratio is given by \cite{byrd_lifetime_2001}
\begin{equation}
    R =  \frac{\int{\rho_{0}^2(\tau)\mathrm{d}\tau}}{\int{\rho^2(\tau)\mathrm{d}\tau}} \, ,
    \label{eq:R}
\end{equation}
where $\rho_0$ is the longitudinal line density corresponding to the zero-current bunch (i.e. a Gaussian distribution with $\sigma = \sigma_{\tau}$), and $\rho$ is the longitudinal line density of the stretched bunch.

Instead of the bunch length, it is this ratio $R$ which should be maximized to gain as much as possible from the double rf system. Moreover, this high $R$ factor should actually be achievable in practice, so it is very important to evaluate the beam stability to judge if the \gls{HC} parameters allow it. To do this, we used an optimizer with the following objective function $f$:

\begin{equation}
    f (\psi_2) = 
    \begin{cases}
        - R & \text{if stable} \\
        x & \text{if unstable} 
    \end{cases}
\end{equation}
Where $x$ is an arbitrary positive number, $x=10$ is chosen here as it is nearly twice as much as the maximum $R$ value considered, used as a penalty factor. The $f$ function presents clear discontinuities at the instability thresholds, for this reason the optimizer we choose is based on the COBYLA algorithm \cite{COBYLA}, which is not based on gradient methods and known to work well with non-differentiable objective functions. The resulting $\psi_2$, which minimizes $f$, corresponds to the largest Touschek lifetime ratio $R$ for which the beam is found to be stable. 

\begin{figure*}
    \centering
    \includegraphics[width=1.0\textwidth]{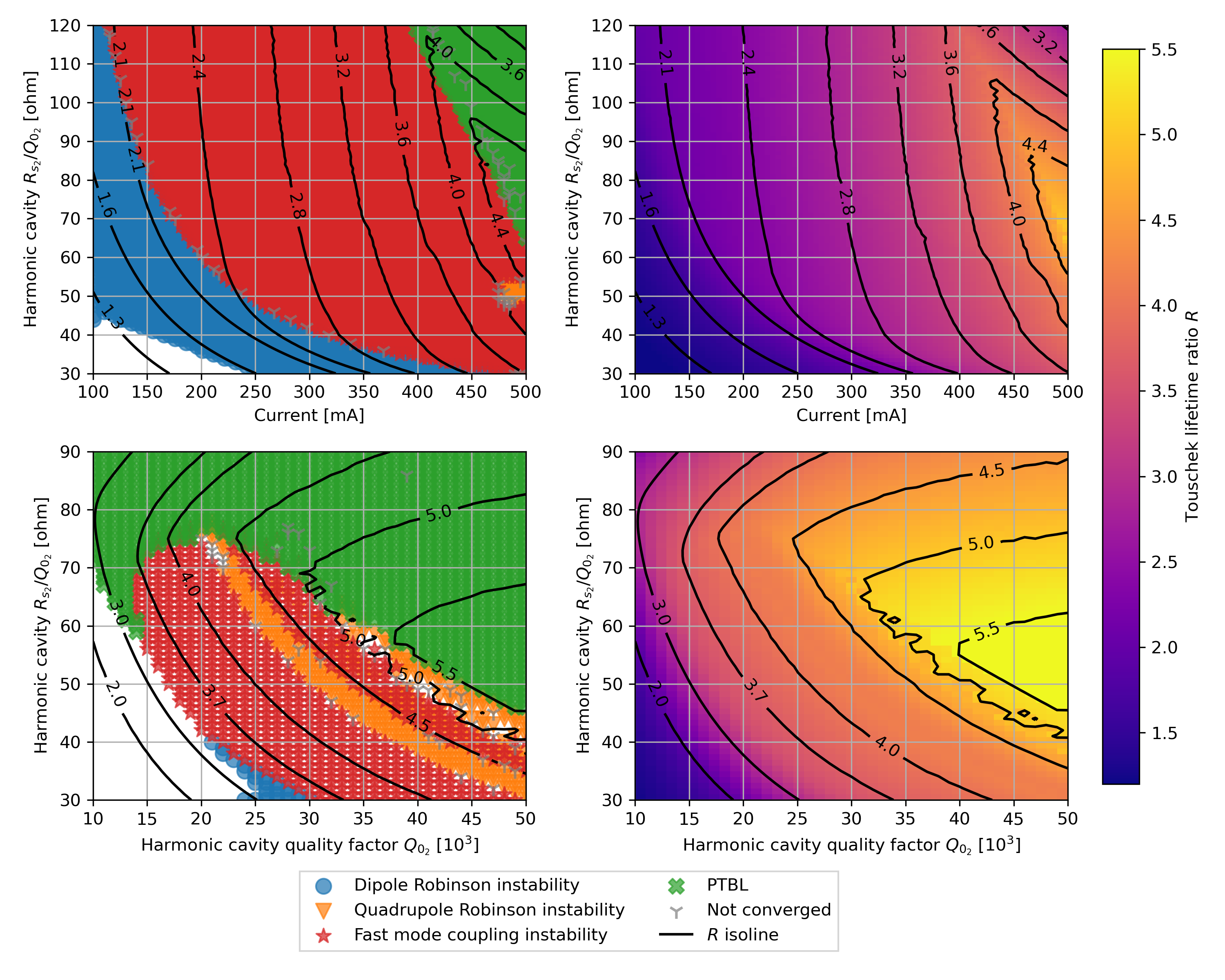}
    \caption{Right: Maximum Touschek lifetime ratio $R$ achievable with a stable beam. Left: Instabilities predicted by a shift of $\Delta \psi_2=$ \SI{-0.1}{\degree} from the $\psi_2$ value giving the maximum Touschek lifetime ratio $R$. Top: $R_{s_2}/Q_{0_2}$ is varied for a constant value of $Q_{0_2} =$ \num{31e3} for different beam current $I_0$. Bottom: $R_{s_2}/Q_{0_2}$ is varied for different values of $Q_{0_2}$ for a beam current of $I_0=$ \SI{500}{\milli \ampere}. The \enquote{Alves} solver is used.}
    \label{fig:optimize}
\end{figure*}

Figure \ref{fig:optimize} shows that this method can be used to find the optimal $R_{s_2}/Q_{0_2}$ for a given $Q_{0_2}$ value for a large beam current $I_0$ range (top) or the optimal value for the pair $(R_{s_2}/Q_{0_2}, Q_{0_2})$ for a given beam current of $I_0= \SI{500}{\milli \ampere}$ (bottom).

The plots on the right of Fig.~\ref{fig:optimize} show the maximum $R$ achievable with a stable beam. To observe which instabilities are limiting to reach higher $R$ factors, the left plot of Fig.~\ref{fig:optimize} shows the instabilities predicted by a shift of $\Delta \psi_2=$ \SI{-0.1}{\degree} from the $\psi_2$ value giving the maximum $R$ (corresponding to the $\psi_2$ values used to get the right plot).

On the top plots, different regions can be observed: when both the beam current $I_0$ and the $R_{s_2}/Q_{0_2}$ are low (where the $R$ isolines are curved), the Touschek lifetime ratio $R$ is low due to the lack of shunt impedance triggering dipole Robinson instabilities before reaching $\xi$ values close to one. When the beam current $I_0$ is low but the $R_{s_2}/Q_{0_2}$ values are higher (where the $R$ isolines are vertical) the limiting instability is the fast mode coupling. In that case, the Touschek lifetime ratio $R$ only depends on beam current $I_0$ and not on the total shunt impedance $R_{s_2}$. At higher currents, the $R$ isolines are looping around optimal regions for the Touschek lifetime corresponding to moderate $R_{s_2}/Q_{0_2}$ values. In this region, the \gls{PTBL} instability is the limiting instability, for which it is known that lower $R_{s_2}/Q_{0_2}$ is favorable \cite{he_periodic_2022}.

The bottom plots show that for a constant total \gls{HC} shunt impedance $R_{s_2}$, it is better to have a lower $R_{s_2}/Q_{0_2}$ and higher $Q_{0_2}$ to get a larger Touschek lifetime. In this case, at \SI{500}{\milli \ampere}, this corresponds to optimizing the \gls{HC} to avoid the \gls{PTBL} instability while maintaining the total \gls{HC} shunt impedance $R_{s_2}$ to be able to obtain the required \gls{HC} voltage. If $R_{s_2}$ gets too low, quadrupole Robinson instability and fast mode coupling instability are triggered and limit further increase of $R$.

\begin{figure}[htbp]
    \centering
    \includegraphics[width=1.0\columnwidth]{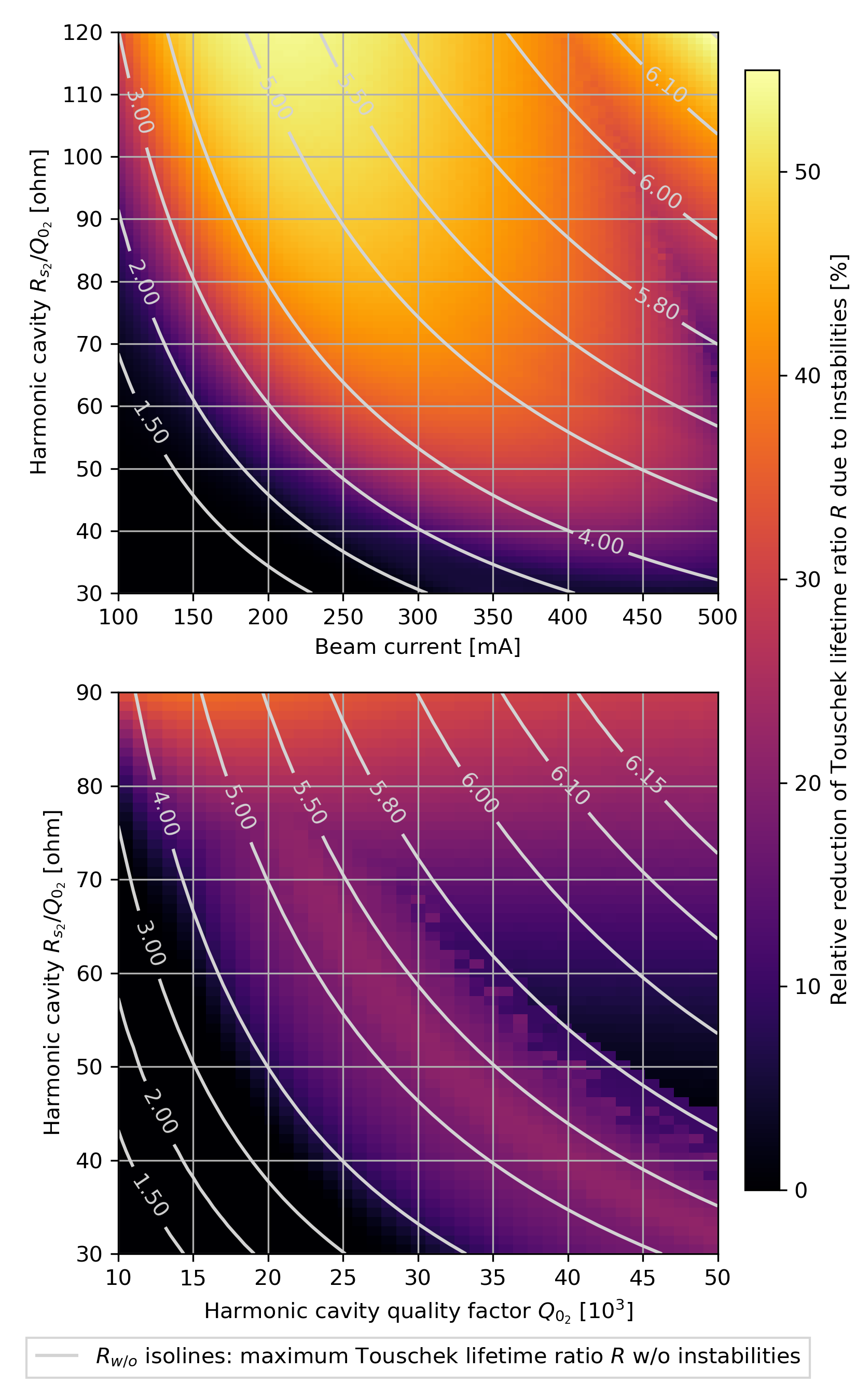}
    \caption{Relative reduction of the Touschek lifetime ratio $R$ due to instabilities and $R_\text{w/o}$ isolines. Top: $R_{s_2}/Q_{0_2}$ is varied for a constant value of $Q_{0_2} =$ \num{31e3} for different beam current $I_0$. Bottom: $R_{s_2}/Q_{0_2}$ is varied for different values of $Q_{0_2}$ for a beam current of $I_0=$ \SI{500}{\milli \ampere}. The \enquote{Alves} solver is used.}
    \label{fig:optimize_eq}
\end{figure}

The regions of the left plots where no instability is triggered by a shift of $\Delta \psi_2=$ \SI{-0.1}{\degree} correspond to an area of the parameter space where the product $R_{s_2} I_0$ is too low to lengthen the bunches effectively. When the product $R_{s_2} I_0$ is too low, the \gls{HC} has to be tuned very close to the resonance to reach the desired voltage. But, close to resonance the \gls{HC} impedance is mostly resistive (real) which is not effective to lengthen the bunches effectively (as bunch lengthening is obtained because of imaginary impedance). Another way to explain it is to say that we can reach the desired \gls{HC} voltage $V_2$ but not the correct phase $\theta_2$, see Fig.~\ref{fig:compare_HC} of Appendix \ref{subsec:MC_settings}. Thus, even without instabilities, a lower $R$ is already expected for low $R_{s_2} I_0$ values, as shown later by the isolines of Fig.~\ref{fig:optimize_eq}.

Using the \enquote{Venturini} solver, the full optimization ran in a bit more than \SI{1}{\hour} on a laptop with the serial version and in a few minutes in the parallelized version. The more accurate optimization using the \enquote{Alves} solver shown here took half an hour in the parallelized version. The results using both solvers are extremely similar for the $(R_{s_2}/Q_{0_2}, I_0)$ optimization but different for the $(R_{s_2}/Q_{0_2}, Q_{0_2})$ case, even if the general structures are the same. The difference is caused by the He criterion used for the \enquote{Venturini} solver for the \gls{PTBL}, which as shown in Sec.~\ref{sec:ptbl}, predicts fewer unstable cases. These very large parameter scans and systematic optimizations are currently out of reach for state-of-the-art tracking codes\footnote{Around 100 instability evaluations are needed for each point in Fig.~\ref{fig:optimize}. So to produce the smallest figure (bottom one) with the same resolution, \num{240e3} instability evaluations are needed in total. Using the same parallelized setup for tracking as in Table~\ref{tab:speed}, it would take more than 250 years.}.

Interestingly, the parameters found to maximize the Touschek lifetime ratio $R$ are very different from the ones computed using the \gls{FP} condition method as shown in Table \ref{tab:opti_param}. At \SI{500}{\milli\ampere}, an \gls{HC} shunt impedance of $R_{s_2} \approx$ \SI{5.65}{\mega \ohm} is needed to reach \gls{FP} conditions as explained in Appendix \ref{subsec:beam_induced}. For a $Q_{0_2} =$ \num{50e3}, the most favorable value shown in the bottom plot of Fig.~\ref{fig:optimize} is $R_{s_2}/Q_{0_2} =$ \SI{113}{\ohm}, which would lead to no more than $R \approx 3.5$. This value is to be compared with the optimal found here for the same $Q_{0_2}$, a $R_{s_2}/Q_{0_2} \approx $ \SI{49}{\ohm} would give a Touschek lifetime ratio close to $R \approx 5.8$.

\begin{table}
        \centering
        \caption{\gls{HC} parameters and performance obtained from the \gls{FP} condition method and the optimizer. Results for the parameters $I_0 = \SI{500}{\milli\ampere}$ and $Q_{0_2} =$ \num{50e3}.}
        \begin{ruledtabular}
        \begin{tabular}{ l c c} 
        {Parameter} & {FP condition} & {Optimizer} \\
        \hline
        $R_{s_2}$ & \SI{5.65}{\mega\Omega} & \SI{2.45}{\mega\Omega}\\
        $R_{s_2}/Q_{0_2}$ & \SI{113}{\Omega} & \SI{49}{\Omega} \\
        $R$ factor & 3.5 & 5.8
        \end{tabular}
        \end{ruledtabular}
    \label{tab:opti_param}
\end{table}

The maximum Touschek lifetime ratio without considering instabilities $R_\text{w/o}$ can be found by using the same optimizer with a different objective function: $g (\psi_2) = - R$, without condition on beam stability. The $R_\text{w/o}$ isolines are shown in Fig.~\ref{fig:optimize_eq} for the same two previous cases. The relative reduction of the Touschek lifetime ratio $R$ due to instabilities, defined as $(R_\text{w/o}-R)/R_\text{w/o}$, is shown in Fig.~\ref{fig:optimize_eq}.

Figure~\ref{fig:optimize_eq} highlights that the optima found in Fig.~\ref{fig:optimize} only happen due to the contribution of different instabilities. 
The local maxima regions of Fig.~\ref{fig:optimize} correspond to a maximum $R_\text{w/o}$, shown here to match with high $R_{s_2} I_0$ values, where the instabilities are the least restrictive (i.e. relative reduction of $R$ due to instabilities is low). 

For SOLEIL~II, the \gls{HC} needs to be used from \SI{200}{\milli\ampere} up to \SI{500}{\milli\ampere} to cover operation in both timing mode and uniform filling mode. Using Fig.~\ref{fig:optimize}, we were able to specify optimal $R_{s_2}/Q_{0_2}$ and $Q_{0_2}$ values for cavity design. Using 2 or 3 \glspl{HC} of $R_{s_2}/Q_{0_2} =$ \SI{30}{\ohm} each with $Q_{0_2} =$ \num{31e3}, to be able to vary the total $R_{s_2}/Q_{0_2}$ from \SIrange{60}{90}{\ohm} is now the leading option for the SOLEIL~II storage ring \gls{HC} system. It is close to the optimal solution and the design seems to be mechanically viable \cite{ESRF_HarmonLIP2024}.

\section{Discussion}
\label{sec:discussion}

\subsection{Cavities and machine impedance short-range wakes}
\label{subsec:non_uniform}

As demonstrated in this study, the existing theories are capable of accurately explaining and predicting the beam dynamics in a double rf system, with results that are comparable to those obtained from tracking simulations. However, this is typically limited to relatively simple cases where the assumptions used to develop the theories are upheld, such as uniform filling patterns and low currents per bunch, which allows for the short-range wakes to be considered negligible.

Two sources of short-range wakes can be distinguished here, from the cavities themselves and from the global machine impedance. It has recently been shown that the beam dynamics can be significantly altered by these wakes when the current per bunch is high. This results in changes to both the achievable bunch length and the instability thresholds \cite{gamelin_beam_2024, He_HarmonLIP2024, Carver_HarmonLIP2024}.

To demonstrate the different effect of these two types of short-range wakes, we consider two cases. Firstly, a 32 bunch mode where the current per bunch is high, but without the machine impedance, to observe the effect of the cavity short-range wakes. This 32 bunch mode corresponds to a filling pattern where the 32 bunches are uniformly distributed in the ring. And, secondly, the same uniform filling mode as previously considered but with the machine impedance.

\begin{figure}
    \centering
    \includegraphics[width=1.0\columnwidth]{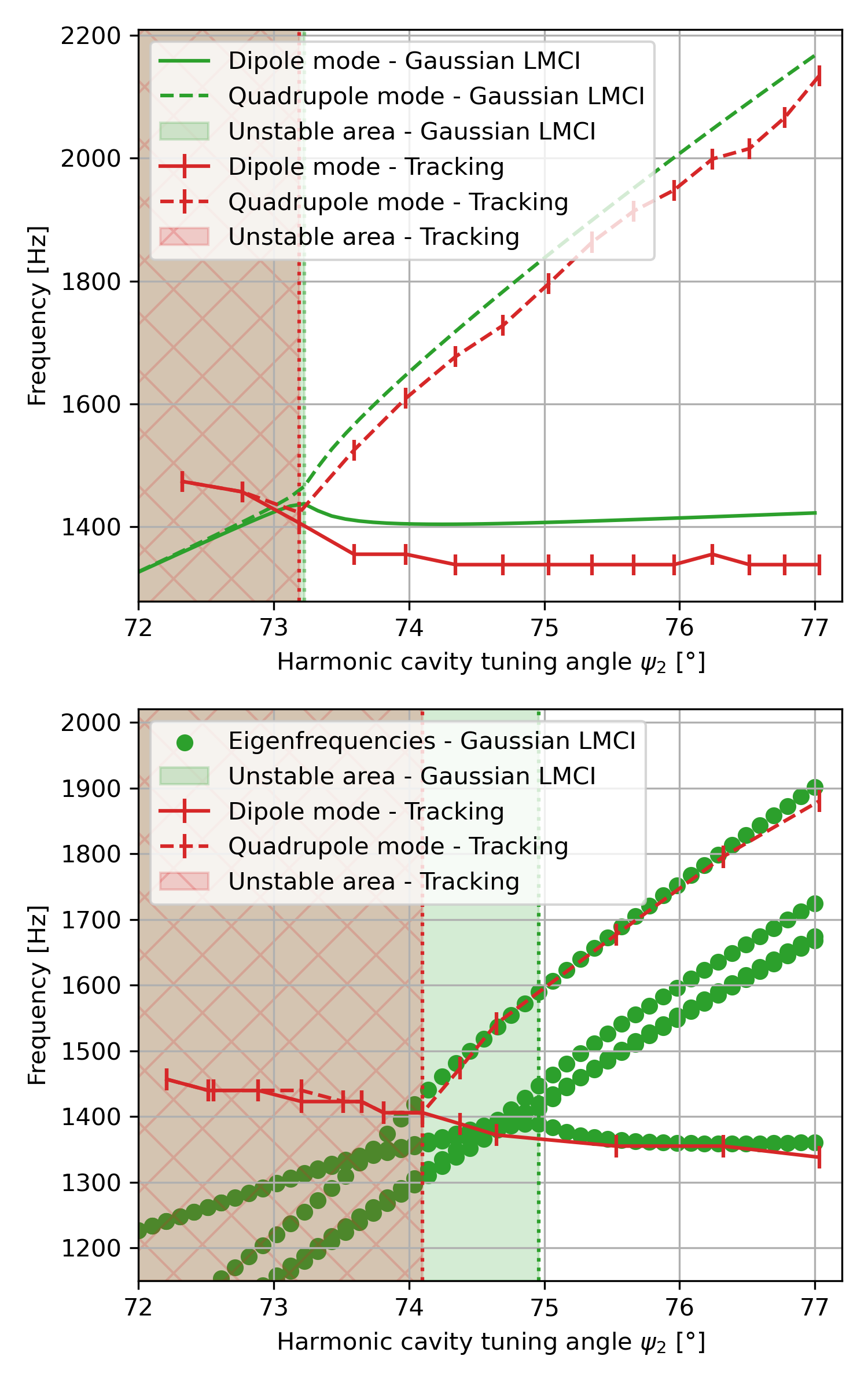}
    \caption{Mode frequencies as function of the \gls{HC} tuning angle $\psi_2$ in the SOLEIL~II storage ring at \SI{200}{\milli \ampere} beam current. Top: 32 bunch filling mode. Bottom: uniform filling mode with impedance model. Passive \gls{HC} parameters: $R_{s_2}/Q_{0_2}=\SI{90}{\ohm}$, $Q_{0_2} =\num{36e3}$.}
    \label{fig:mode_coupling_disscussion}
\end{figure}

Figure~\ref{fig:mode_coupling_disscussion} illustrates this effect with the fast mode coupling in the same conditions as Fig.~\ref{fig:mode_coupling} but for 32 bunch filling (top) and for uniform filling with SOLEIL~II storage ring impedance model (bottom).
The impedance model of SOLEIL~II includes contributions from over 50 distinct elements with main contributions coming from resistive wall and tapers \cite{Gubaidulin_SOLEILII_2024}.

The tracking data is compared with analytical computations using the Gaussian LMCI method with the \enquote{Alves} solver. This allows for including the cavities and impedance model short-range wakes in both the Ha\"{i}ssinski equation and the instability calculation. The Bosch theory is unable to treat these cases, at least without significant modifications, and thus is not represented here. When the SOLEIL~II storage ring impedance model is incorporated into the tracking process, \num{100000} macro-particles per bunch are utilized instead of \num{10000} in order to accurately model the high-frequency short-range wakes.

In 32 bunch mode, the fast mode coupling instability is observed to happen at a lower \gls{HC} tuning angle $\psi_2$ than in the uniform filling case of Fig.~\ref{fig:mode_coupling}. We can see from the tracking results that the quadrupole mode is shifted in the upward direction: at $\psi_2 =$ \SI{77}{\degree} the quadrupole frequency is \SI{1965}{\hertz} in uniform filling and \SI{2134}{\hertz} in 32 bunch mode while the dipole frequency remains unchanged. Physically, it can be explained by the stronger short-range cavity wakes (proportional to bunch current) in 32-bunch mode. The short-range cavity wakes, both from \gls{MC} and \gls{HC}, are capacitive in nature, so they will tend to shorten the bunch length and induce a positive incoherent tune shift. This tune shift is compensated by the dynamic coherent frequency shift\footnote{The dynamic coherent frequency shift results from the fact that, for a dipole motion, the short-range wake moves with the bunch.} for the dipole mode but not for the quadrupole mode \cite{[{See pp. 68 and pp. 206 of }]ng_physics_2006}. This effect is reproduced with a good agreement by the Gaussian LMCI method.

When the machine impedance model is included in uniform filling mode, the same effect is observed but in the opposite direction. The machine impedance is mostly inductive, leading to a negative incoherent tune shift which is only observed in tracking on the quadrupole mode frequency. At $\psi_2 =$ \SI{77}{\degree}, the quadrupole frequency is \SI{1880}{\hertz} with the impedance model included and \SI{1965}{\hertz} without, whereas the dipole frequency is mostly unchanged. To reproduce this effect using the Gaussian LMCI method, radial modes up to $k_{max}=3$ are required to be included in the analysis compared to the case presented previously where only the first radial mode associated with the dipole $m=1$ and quadrupole azimuthal modes $m=2$ were sufficient. Even then, only the eigenfrequencies of the modes fit the tracking data. The Gaussian LMCI method predicts an instability at $\psi_2 \approx$ \SI{75}{\degree} compared to $\psi_2 \approx$ \SI{74}{\degree} with tracking, probably from mode coupling between two radial modes but this is not observed in tracking.

These cases show how the combination of the Ha\"{i}ssinski equation and Gaussian LMCI theory can be used to explain more complex cases including machine impedance and short-range cavity wakes, although the agreement is not perfect with the detailed particle tracking simulations.

\subsection{Limitations and possible improvements}

We have shown that the Gaussian LMCI method is quite successful in predicting the fast mode coupling instability. In fact, the Gaussian LMCI method is also able to estimate other instabilities of coupled bunch mode $\ell=0$, such as the dipole and quadrupole Robinson instabilities, if sufficient order of radial modes are included. The issue is, when many azimuthal $m$ and radial $k$ modes are included, the mode classification, and thus the instability identification, is much more difficult to do in an automated way.

Bosch theory is less general but provides different expressions for each instability type, thus the instability identification is much simpler. For algorithms such as the ones developed in this paper, instability identification is as important as knowing if there is an instability or not. For example, when designing a system it is crucial to differentiate a dipole Robinson instability from a quadrupole Robinson instability since the dipole instability can be expected to be damped by a longitudinal dipolar bunch-by-bunch feedback but not the quadrupolar one.

For the same reasons, it is favorable that semi-analytical methods tend to yield more unstable solutions than tracking simulations. This is to be expected because some unstable solutions with low growth rates will not be observed in tracking. Consequently, we are not missing any potentially unstable solution.

The modified algorithms have been designed and benchmarked mainly for \gls{NC} passive \glspl{HC}. They should, in principle, also work for \gls{NC} active \glspl{HC}, as is the case in Bosch's original algorithm, but this has not yet been benchmarked. For \gls{SC} passive \glspl{HC}, the modified algorithms are also expected to perform well as the physics is mostly very similar to \gls{NC} passive \glspl{HC}. 

A major difference between \gls{NC} passive \glspl{HC} and \gls{SC} passive \glspl{HC} is that for the latter, the \gls{HC} detuning $\Delta \omega = \omega_r - \nu \omega_{\mathrm{rf}}$ can be very small due to its very high shunt impedance. In that case, an instability can be driven by the mode coupling between the dipole mode and the cavity mode.
The cavity mode mostly follows $\omega = \Delta \omega$ and originates from the beam modulation at $\omega_{\mathrm{rf}} \pm \Delta \omega$ feeding back into the cavity \cite{towne_spectrum_1998}. This instability has been recently studied in the context of the operation of \gls{SC} passive \gls{HC} at low total current \cite{he_mode-zero_2023, gamelin_beam_2024} and it would need to be added in the algorithm to correctly estimate \gls{SC} passive \glspl{HC} beam dynamics in this particular regime.

Another improvement that could be made is to use a better Landau damping model. As shown in Sec.~\ref{sec:res}, the Landau damping estimate from Bosch's model can be rather pessimistic compared to tracking when $\xi \approx 1$. It is not an issue for instabilities whose mechanism is mode coupling of some sort, like the \gls{PTBL} or the fast mode coupling instability, as their growth rates tend to be strong. But for instabilities originating from a single mode, like \gls{CBI} driven by a \gls{HOM} or even dipole and quadrupole Robinson instabilities, whose growth rates tend to increase gradually with intensity and can be potentially Landau damped, there is possibly an accuracy issue near $\xi=1$.

\section{Conclusion}

This study presents modernized semi-analytical algorithms to predict longitudinal beam instabilities in double rf systems of synchrotron light sources. By incorporating recent advances in the theory, the new algorithms consider not only Robinson instabilities but also the \gls{PTBL} as an additional instability, allowing for better precision and better convergence against numerical parameters. The new algorithms have been benchmarked with excellent results against multibunch tracking, but are orders of magnitude faster. The re-implementation of the original Bosch algorithm and the new algorithms are distributed as an open-source python tool \gls{ALBuMS}, which relies on the collective effects libraries \texttt{pycolleff} and \texttt{mbtrack2}.

In addition, \gls{ALBuMS} has been used to do parameter scans and optimization of the \gls{HC} in a very large parameter space to maximize the achievable Touschek lifetime in the framework of the SOLEIL~II project. It has been found that the optimal \gls{HC} parameters are far from the one leading to the \gls{FP} conditions, leading to a 65\% larger Touschek lifetime when optimized. We show that the different instability contributions fully determine the optimum. Optimization studies of this scale are only possible using the type of semi-analytical algorithms developed in this article, and can drastically simplify dual rf system design for future light sources.

Finally, to illustrate further studies and limitations in the algorithm, we show how more complex cases with cavity and machine impedance short-range wakes can be partially included in these methods and demonstrate their impact on the fast mode coupling instability.

\begin{acknowledgments}
The authors thank the rf and accelerator physics groups of the SOLEIL synchrotron and Patrick Marchand for useful discussions.
The authors thank F.~H.~de~S{\'a} for proofreading the manuscript.
Part of this work was performed using CCRT HPC resource (TOPAZE supercomputer) hosted at Bruyères-le-Châtel, France.
This article is dedicated to the memory of Jacques Ha\"{i}ssinski.
\end{acknowledgments}

\appendix

\section{Double rf systems}
\label{sec:double_rf_sys}
\subsection{Flat potential conditions}

Let us assume an rf system composed of two cavities: a \gls{MC} with a voltage $V_1$, a phase $\theta_1$ working at the angular rf frequency $\omega_{\mathrm{rf}}$ and a \gls{HC} of the $\nu$\textsuperscript{th} harmonic with a voltage $V_2$ and phase $\theta_2$. The overall voltage $V_{tot}$ provided by this system at time $t$ is given by:
\begin{equation}
    V_{tot} (t) = V_1 \cos{(\omega_{\mathrm{rf}} t + \theta_1)} + V_2 \cos{(\nu \omega_{\mathrm{rf}} t + \theta_2)} \, .
\end{equation}
The energy loss per turn $U_{0}$ must be compensated by the total rf voltage ${V}_{tot}$ for the synchronous particle, i.e. $eV_{tot} (0) = U_{0}$, where $e>0$ is the elementary charge. Then the \gls{MC} phase $\theta_1$ is given by:
\begin{equation}
    \cos{\theta_1} = \frac{U_0}{e V_1} - \frac{V_2}{V_1}\cos{\theta_2} \, .
    \label{eq:theta1_noxi}
\end{equation}
In such a system, assuming small oscillations (i.e. $\tau \ll 1$), the linear synchrotron angular frequency $\omega_{s}$ can be written as:
\begin{equation}
    \omega_{s}^2 = \frac{e \eta \omega_{\mathrm{rf}}}{E_0 T_0} (1 - \xi) V_1 \sin{\theta_1} = \omega_{s_0}^2 (1 - \xi)
\end{equation}
Where $\eta$ is the slip factor, $E_0$ is the storage ring reference energy, $T_0$ is the revolution time, $\omega_{s_0}$ is the linear synchrotron angular frequency in single rf systems and $\xi$, the ratio of the harmonic to main cavity \enquote{force}, is introduced as defined in \cite{bosch_robinson_2001}:
\begin{equation}
    \xi = \frac{- \nu V_2 \sin{\theta_2}}{V_1 \sin{\theta_1}} \, .
    \label{eq:xi_bosch}
\end{equation}
One can adjust the bunch length by controlling the slope of the overall voltage at the synchronous phase, $\dot{V}_{tot} (0)$, and its derivative, $\ddot{V}_{tot} (0)$. If $eV_{tot} (0) = U_{0}$, together with $\dot{V}_{tot} (0) = \ddot{V}_{tot} (0) = 0$ then the system is said to be at the \acrfull{FP} condition.

\subsection{Near flat potential conditions}
\label{sec:nfp}

For a passive \gls{HC}, the voltage and the phase of the \gls{HC} can not be set independently. In that case, for a fixed $V_1$, $\theta_1$ and $V_2$ are the only free parameters left to set $V_{tot} (0)$ and $\dot{V}_{tot} (0)$, $\ddot{V}_{tot} (0)$ can not be controlled. Expressing $\theta_1$ and $V_2$ by introducing the $\xi$ variable gives:
\begin{equation}
    \cos{\theta_1} = \frac{U_{0}}{e V_1} + \xi \frac{\sin{\theta_1}}{\nu \tan{\theta_2}} \, ,
    \label{eq:balance_xi}
\end{equation}
\begin{equation}
    V_2 = - \xi \frac{V_1 \sin{\theta_1}}{\nu \sin{\theta_2}} \, ,
    \label{eq:V2_xi}
\end{equation}
For a fixed $\xi$, Eq. \eqref{eq:balance_xi} is a self-consistent equation for $\theta_1$, and the existence of a solution depends on the value of $\theta_2$. Now expressing the first and second derivatives of the rf voltage gives:
\begin{equation}
    \dot{V}_{tot} (0) = \xi \omega_{\mathrm{rf}} \sin{\phi_1} \left ( 1 - \frac{1}{\xi} \right) \, .
    \label{eq:alpha_xi}
\end{equation}
\begin{equation}
    \ddot{V}_{tot} (0) = - V_1 \omega_{\mathrm{rf}}^2 \left [ \cos{\theta_1} - \xi \nu \frac{\sin{\theta_1}}{\tan{\theta_2}} \right] \, .
    \label{eq:alpha2_xi}
\end{equation}
One can see that the value $\xi=1$ cancels the slope of the rf voltage $\dot{V}_{tot} (0)$ but not necessarily its derivative $\ddot{V}_{tot} (0)$. To get the particular case of $\ddot{V}_{tot} (0) = 0$, another condition on $\theta_2$ needs to be met:
\begin{equation}
    \tan{\theta_2} = \xi \nu \tan{\theta_1}
    \label{eq:phi2_xi}
\end{equation}
To distinguish both cases, we will call the condition $\xi=1$ in Eq. \eqref{eq:balance_xi} and \eqref{eq:V2_xi}, corresponding to $\dot{V}_{tot} (0) = 0$ and $\ddot{V}_{tot} (0) \neq 0$, the \acrfull{NFP} conditions. While the \gls{FP} condition corresponds to $\xi=1$ in Eq. \eqref{eq:balance_xi}, \eqref{eq:V2_xi} and \eqref{eq:phi2_xi}. The main consequence of a non-zero $\ddot{V}_{tot} (0)$ is an asymmetry in the bunch distribution.

\subsection{Beam-induced voltage}
\label{subsec:beam_induced}
The beam induced voltage $V_b$ inside a cavity, assuming a uniform beam filling pattern, is given by \cite{venturini_passive_2018}:
\begin{equation}
    \label{eq:beam_loading_voltage}
    V_{b} (t) = - \frac{2 I_0 R_s F}{1 + \beta} \cos{(\psi)} \cos{(\nu \omega_{\mathrm{rf}} t +  \psi - \Phi)} \, .
\end{equation}
Where $I_0$ is the total beam current, $R_{s}$ is the cavity shunt impedance and $\beta$ is the cavity coupling. The beam-induced voltage $V_b$ is controlled by changing the cavity resonance frequency $\omega_r$ (usually by using a mechanical tuner). The cavity tuning angle $\psi$ is defined as:
\begin{equation}
    \tan{\psi} = Q_L \left ( \frac{\omega_r}{\nu \omega_{\mathrm{rf}}} - \frac{\nu \omega_{\mathrm{rf}}}{\omega_r} \right ) \, ,
    \label{eq:tuning_angle}
\end{equation}
where $Q_L$ is the cavity loaded quality factor and $\nu$ is the harmonic of the cavity.

Both voltage and phase are further modified by the complex bunch form factor $F e^{i \Phi} = \mathcal{F}[\rho] (\nu \omega_{\mathrm{rf}})$ which depends on the Fourier Transform ($\mathcal{F}$) of the bunch profile $\rho$ at the $\nu^{th}$ harmonic:
\begin{equation}
    F e^{i \Phi} = \mathcal{F}[\rho] (\nu \omega_{\mathrm{rf}}) = \int_{-\infty}^{\infty} {e^{i \nu \omega_{\mathrm{rf}} t} \rho (t) dt}
    \label{eq:form_factor}
\end{equation}
For passive \gls{HC}, the total cavity voltage is equal to the beam-induced voltage. Without a power coupler, $\beta_2=0$ and $Q_{L_2} = Q_{0_2}$, and the \gls{HC} voltage $V_2$ and phase $\theta_2$ are expressed as:
\begin{equation}
    \begin{split}
        & V_2 = -2 I_0 R_{s_2} F_2 \cos{\psi_2} \\
        & \theta_2 = \psi_2 - \Phi_2
    \end{split}
    \label{eq:passive_Vp}
\end{equation}
For an active cavity, the total cavity voltage is the vector sum of both the beam-induced voltage component and the generator voltage component.

If we assume that the bunch form factor is purely real ($\Phi=0$), which is true for a quartic potential, then the \gls{FP} condition can only be met for a single value of the $R_{s_2} I_0$ product: 
\begin{equation}
    R_{s_2} I_0 = \frac{ V_1 \cos{\theta_1}}{2 \nu^2 \cos^2{\psi_2} F_2} \, .
\end{equation}
For SOLEIL~II parameters and a beam current of $I_0 =$ \SI{500}{\milli\ampere}, it gives\footnote{Assuming a \gls{std} bunch length $\sigma_\tau=$ \SI{40}{\pico \second} and a Gaussian bunch shape with real form factor $F_2 = e^{-1/2 (\nu \omega_{\mathrm{rf}} \sigma_\tau)^2} \approx 0.94$.} a shunt impedance of $R_{s_2} \approx$ \SI{5.65}{\mega \ohm}. For a more precise determination of the shunt impedance needed to reach \gls{FP} condition, one needs to solve the previous equations in a self-consistent way \cite{venturini_passive_2018, tavares_equilibrium_2014}.

\subsection{Main cavity settings}
\label{subsec:MC_settings}

Depending on the \gls{HC} type (active or passive, \gls{NC} or \gls{SC}), the \gls{MC} will have to be set differently: 

For a \gls{NC} passive \gls{HC}, the \gls{MC} total phase $\theta_1$ needs to be set to cancel the losses from the \gls{HC} (in addition to the synchrotron radiation energy loss per turn $U_0$). Neglecting form factors ($F_2=1$ and $\Phi_2=0$) and introducing Eq. \eqref{eq:passive_Vp} in Eq. \eqref{eq:theta1_noxi}, it gives:
\begin{equation}
    \cos{\theta_1} = \frac{U_0}{e V_1} + \frac{2 I_0 R_{s_2} \cos^2{\psi_2}}{e V_1} \, .
    \label{eq:theta1_HC}
\end{equation}

For a \gls{SC} passive \gls{HC}, $\psi_2 \approx \pi/2$ for most of the useful tuning range. So, in first approximation the losses in the \gls{HC} are negligible and $\theta_1$ is the same as the one used without \gls{HC}, i.e. $\cos{\theta_1} = \frac{U_0}{e V_1}$.

For an active \gls{HC}, the \gls{MC} phase is fixed at its value for \gls{FP} conditions:
\begin{equation}
    \cos{\theta_1} = \frac{\nu^2}{\nu^2-1} \frac{U_{0}}{e V_1} \, ,
\end{equation}
and it is the $\theta_2$ which is actively varied to cancel $\ddot{V}_{tot} (0)$.

\begin{figure}
    \centering
    \includegraphics[width=1.0\columnwidth]{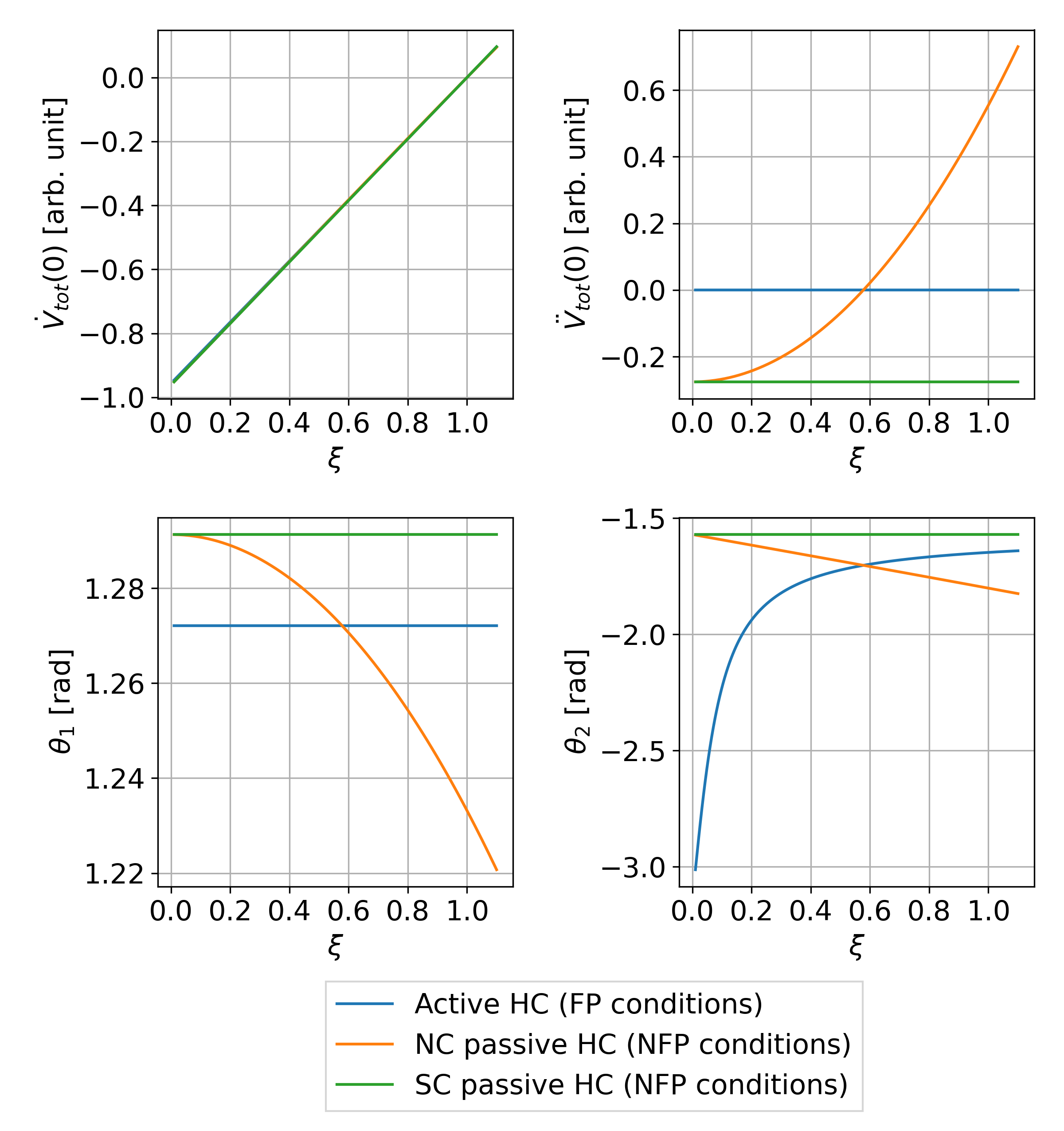}
    \caption{Evolution of the first and second derivative of the total rf voltage $\dot{V}_{tot} (0)$ and $\ddot{V}_{tot} (0)$, of the \gls{MC} and \gls{HC} phase $\theta_1$ and $\theta_2$ versus the $\xi$ parameter. Values used for the \gls{NC} passive \gls{HC} case: $I_0 =$ \SI{500}{\milli\ampere}, $R_{s_2} =$ \SI{60}{\ohm} $\times$ \num{31e3} = \SI{1.86}{\mega \ohm}.}
    \label{fig:compare_HC}
\end{figure}

The different cases are compared in Fig. \ref{fig:compare_HC}. In this figure, Eq. \eqref{eq:balance_xi} and \eqref{eq:V2_xi} are solved self-consistently for different $\xi$ values to show the rf parameter evolution for different \gls{HC} types. In this simple model, both the \gls{MC} and \gls{HC} form factors are neglected ($F_1=F_2=1$ and $\Phi_1=\Phi_2=0$) and the difference between the \gls{HC} type is simply given by how $\theta_2$ is set: using Eq. \eqref{eq:phi2_xi} for the active \gls{HC}, $\theta_2 \approx - \pi / 2$ for \gls{SC} passive \gls{HC} and $\theta_2 = \arccos(\xi V_1/(2 \nu I_0 R_{s_2} ))$ for \gls{NC} passive \gls{HC}.
It can be seen that in the \gls{NC} passive case, the second derivative of the rf voltage $\ddot{V}_{tot} (0)$ is only canceled for a single value of $\xi$ which is well below $\xi=1$. The reason is that for $I_0 =$ \SI{500}{\milli\ampere}, the chosen \gls{HC} shunt impedance $R_{s_2} =$ \SI{60}{\ohm} $\times$ \num{31e3} = \SI{1.86}{\mega \ohm} is well bellow the value needed to get to the \gls{FP} condition found earlier.

Usually \gls{MC} are operated at (or close to) the optimum cavity tuning, which minimizes the forward power from the generator and the reflected power from the cavity. This condition is given by \cite{wilson1994fundamental}: 
\begin{equation}
    \tan{\psi_1} = - \frac{2 I_0 R_{s_1} F_1}{V_1 \sin{\theta_1} (1 + \beta_1)} \, ,
    \label{eq:optimal_tuning}
\end{equation}
where, unless known in advance, $F_1 = 1$ is usually assumed.

\section{Convergence study}
\label{sec:convergence}

\begin{figure}
    \centering
    \includegraphics[width=1.0\columnwidth]{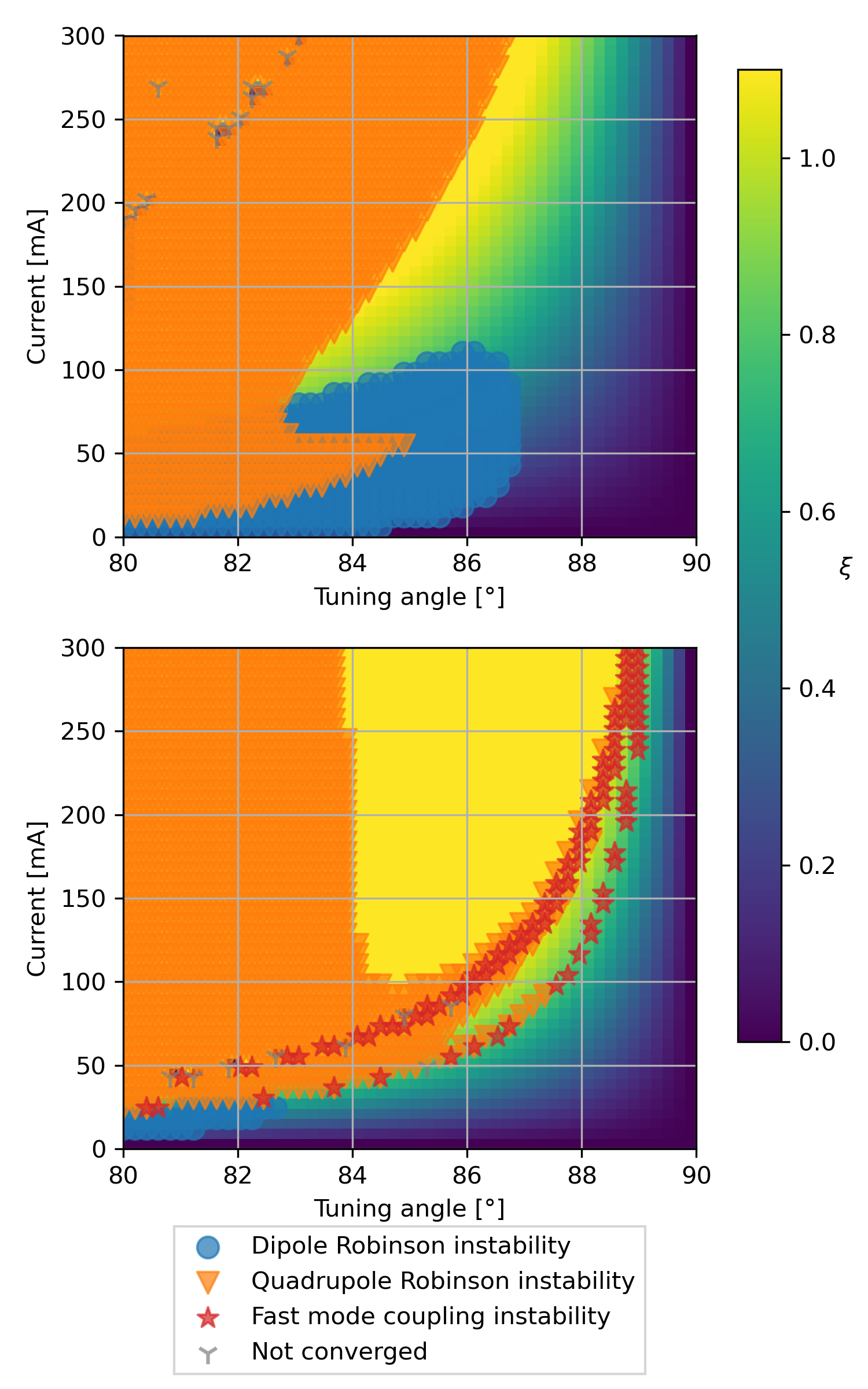}
    \caption{Robinson instabilities predicted for the Aladdin \enquote{standard} lattice (top) and Aladdin \enquote{LF20} lattice (bottom) using the \enquote{Bosch} solver with $\tau_{B} = 20 \sigma_\tau$.}
    \label{fig:bosch_convergence}
\end{figure}

To make sure that the original Bosch algorithm was implemented correctly, the different figures of \cite{bosch_robinson_2001, bosch_instabilities_2005} showing instability prediction in various cases were reproduced. Figure \ref{fig:bosch_convergence} shows the instabilities predicted for the Aladdin \enquote{standard} lattice and Aladdin \enquote{LF20} lattice corresponding to Fig. (1.b) and (5.b) of \cite{bosch_robinson_2001}. The plots are reversed because the definition of the tuning angle $\psi$ used here, Eq. \eqref{eq:tuning_angle}, is opposite of the one used in \cite{bosch_robinson_2001}.

The choice of the integration boundary $\tau_B$ for the bunch length calculation, Eq. (8) in \cite{bosch_robinson_2001}, is a key parameter to be able to reproduce the results. Once the correct choice of integration boundary was found, it was possible to reproduce accurately all the figures, including the active \gls{HC} cases. The full benchmark is joined to the code repository \cite{albums}.

\begin{figure*}[htbp]
    \centering
    \includegraphics[width=1.0\textwidth]{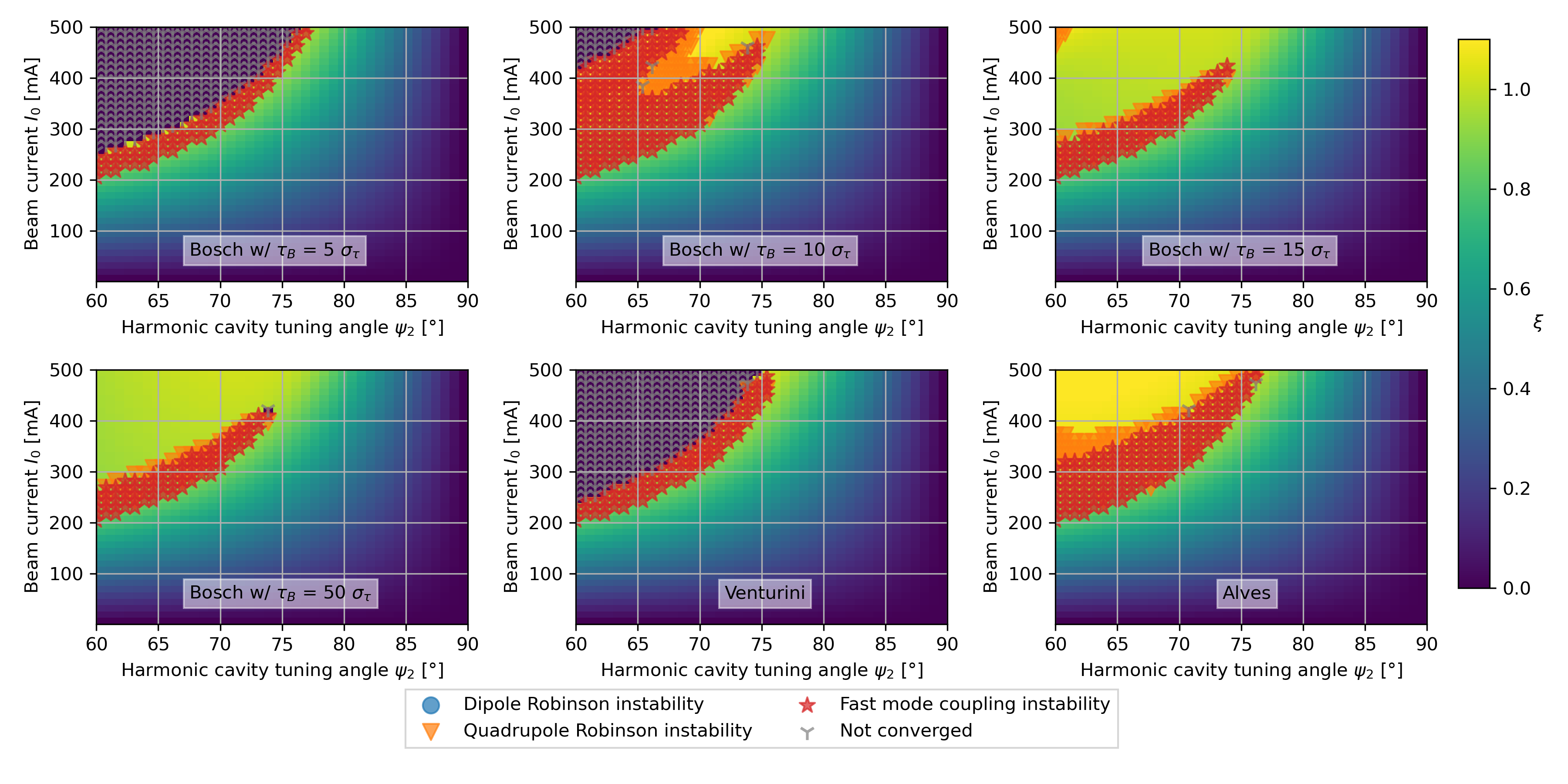}
    \caption{Robinson instabilities predicted for the SOLEIL~II lattice with the \enquote{Bosch} solver with different $\tau_{B}$ values, with the \enquote{Venturini} and \enquote{Alves} solvers. The \gls{PTBL} instability is not shown for solvers \enquote{Venturini} and \enquote{Alves}. Passive \gls{HC} parameters: $R_{s_2}/Q_{0_2}=\SI{60}{\ohm}$, $Q_{0_2} =\num{31e3}$.}
    \label{fig:soleil_convergence}
\end{figure*}

Bosch indicates that \enquote{the integration should be taken over an interval in $\tau$ that greatly exceeds $\sigma_{\tau}$ but is smaller than the confining rf bucket}. The impact of the variation of the integration boundaries $\tau_B$ is shown with SOLEIL~II parameters in Fig.~\ref{fig:soleil_convergence}. For relatively small values, $5 \tau_0 \leq \tau_B \leq 15 \tau_0$, the instability prediction changes nearly continuously with $\tau_B$. For this case, the region where the \enquote{Bosch} solver seems to give consistent results is between $\tau_B= 20 \sigma_\tau$ and $\tau_B= 100 \sigma_\tau$. By consistent, we mean when the results do not vary too much with a small variation of $\tau_B$. Increasing this value further leads to large areas without convergence. The range of $\tau_B$ for which the instability prediction is stabilized depends on the lattice parameters, in particular in the ratio $\sigma_\tau / T_1$. For the Aladdin case, using $\tau_B= 50 \sigma_\tau$ already gives quite different results shown in Fig.~\ref{fig:bosch_convergence}, which are computed for $\tau_B= 20 \sigma_\tau$ in this paper.

To better display the difference between the Robinson instabilities predicted in the different models, the \gls{PTBL} instability is not shown for solvers \enquote{Venturini} and \enquote{Alves} in Fig.~\ref{fig:soleil_convergence}. The modified algorithms allow us to get the instability prediction independent of this numerical parameter.
For the original Bosch algorithm, a convergence study, like the one above, is necessary.

In addition, in the region of $\tau_B$ where the \enquote{Bosch} solver seems to have converged (between $\tau_B= 20 \sigma_\tau$ and $\tau_B= 100 \sigma_\tau$), the original algorithm do not predict a fast mode coupling instability in the area between \SIrange{450}{500}{\milli \ampere} and \SIrange{75}{80}{\degree}, while one is predicted using the \enquote{Venturini} or \enquote{Alves} solvers. In Sec.~\ref{sec:tracking}, we confirmed with tracking simulations that such an instability is indeed expected. Note that the fast mode coupling instability is predicted in that area with the \enquote{Bosch} solver when $\tau_B = 5 \sigma_\tau$, but in that case, the results are extremely dependent on $\tau_B$.

Furthermore, in the region $20 \tau_0 \leq \tau_B \leq 100 \tau_0$, the original algorithm predicts a large stable area past the \gls{NFP} ($\xi>1$) separated by the stable area where $\xi < 1$ by an unstable zone due to the fast mode coupling instability. When using the \enquote{Venturini} or \enquote{Alves} solvers, this region is shown as either unstable due to \gls{PTBL} instability or not converged because no solution to the Ha\"{i}ssinski equation has been found, see Fig.~\ref{fig:HOM_Alves} where the \gls{PTBL} is included. As shown in Fig.~\ref{fig:soleil_convergence}, when using the original algorithm, the equation used for mode $\ell=1$ instability does not predict instabilities there. 

The \enquote{Venturini} solver is less robust than the \enquote{Alves} one to get the bunch profile when $\xi > 1.1$, which explains why there is a large zone of no convergence for this method in this area. The \enquote{Alves} solver robustness relies on employing Anderson's acceleration algorithm to compute the fixed-point solution of Haissinki equation \cite{warnock_equilibrium_2021-1}. In most cases, it should not be an issue as systems in which $\xi > 1.1$ are usually unstable or have two split bunches in the same rf bucket, which is unfavorable for Touschek lifetime or other collective effects. 

Apart from \gls{PTBL}, the three solvers handle the other instabilities considered in the same way. There are sometimes differences in the instability predictions between the different methods despite being based on the same equations. This is a consequence of different results for the self-consistent problem, giving different equilibrium voltages in the cavities and form factors. Even so, all these differences are usually small compared to the large ones shown here for the Robinson instabilities when $\tau_B$ is varied for the \enquote{Bosch} solver.

\bibliography{bib}

\end{document}